\documentclass[a4paper,11pt,reqno]{article}
\usepackage{a4wide}
\usepackage{amsmath,amsfonts,amssymb}
\usepackage[english]{babel}
\usepackage{soul}
\usepackage{nicefrac}
\usepackage[mathscr]{euscript}
\usepackage{setspace}
\usepackage{datetime}
\usepackage[nosort]{cite}
\usepackage{mathrsfs}
\usepackage[T1]{fontenc}
\usepackage{txfonts}
\usepackage{kpfonts}
\usepackage{upgreek}
\usepackage{comment}
\usepackage{mathpazo}
\usepackage{graphicx}
\usepackage[breaklinks=true]{hyperref}
\usepackage{color}
\usepackage{fancybox}

\newcommand{\p}{\partial}

\newcommand{\F}{\Phi}
\newcommand{\sg}{\sqrt{-g}}
\DeclareMathOperator{\extdm}{d}
\newcommand{\extd}{\extdm \!}

\def\beq{\begin{equation}}
\def\eeq{\end{equation}}

\numberwithin{equation}{section}

\hypersetup{colorlinks=true,
			urlcolor=blue,
			citecolor=magenta,
			linkcolor=blue,}

\let\oldsqrt\sqrt
\def\sqrt{\mathpalette\DHLhksqrt}
\def\DHLhksqrt#1#2{%
\setbox0=\hbox{$#1\oldsqrt{#2\,}$}\dimen0=\ht0
\advance\dimen0-0.2\ht0
\setbox2=\hbox{\vrule height\ht0 depth -\dimen0}%
{\box0\lower0.4pt\box2}}

\newcommand{\RNum}[1]{\uppercase\expandafter{\romannumeral #1\relax}}

\author{
  \begin{minipage}{.97\linewidth}
   \vspace{1cm}
       \begin{center}
                \textbf{St\'ephane Detournay}$^{1}$,
      \textbf{P. Marios Petropoulos}$^2$,\\
      \textbf{C\'eline Zwikel}$^{3}$
    \end{center}
    \vspace{0.5cm}
      \hspace{2.4cm}
\begin{center}   
{\it \begin{footnotesize}
1: Physique Math\'ematique des Interactions Fondamentales  \\
	Universit\'e Libre de Bruxelles, Campus Plaine - CP 231, 1050 Bruxelles, Belgium \\
	sdetourn@ulb.ac.be\\
	\vspace{0,5cm}
2: Centre de Physique Th\'eorique \\
	Ecole Polytechnique, CNRS UMR 7644, Universit\'e Paris-Saclay, 91128 Palaiseau Cedex, France\\
	marios.petropoulos@polytechnique.edu\\
	\vspace{0,5cm}
3: Institute for Theoretical Physics\\
Vienna University of Technology, Wiedner Hauptstrasse 8-10/136, 1040 Wien, Austria\\
zwikel@hep.itp.tuwien.ac.at
 		\end{footnotesize}}
\end{center}
    \vspace{0.5cm}
  \end{minipage}
}

\title{\vspace{1.5cm}
 \boldmath 
   \begin{huge}  
    \textbf{Asymptotic Symmetries of Three-Dimensional Black Strings}
  \end{huge} 
  \unboldmath
}

\date{}

\begin{document}

\begin{titlepage}
\maketitle
\thispagestyle{empty}

 \vspace{-14.cm}
  \begin{flushright}
  CPHT-RR079.082018
  \end{flushright}
 \vspace{12.7cm}

\begin{center}
\textsc{Abstract}\\  
\vspace{0.7cm}	
\begin{minipage}{1.0\linewidth}

We determine a consistent phase space for a theory consisting in the Einstein--Hilbert action coupled to matter fields (dilaton, one-form, two-form) and containing three-dimensional black strings (the Horne--Horowitz solution and generalizations thereof). The theory at hand is the low energy effective action for the bosonic sector of heterotic string theory.  We find a consistent set of boundary conditions whose algebra of asymptotic charges consist in a single Virasoro algebra supplemented by three global $u(1)$ generators.  We also discuss the thermodynamics of the zero-mode solutions and point out some peculiar features of this system.

\end{minipage}
\end{center}


\end{titlepage}

\onehalfspace

\begingroup
\tableofcontents
\endgroup
\noindent\rule{\textwidth}{0.6pt}

\section{Introduction}

Three-dimensional gravity has since more than three decades played a very special r\^ole in uncovering the nature of the gravitational interaction beyond the classical level. Its simplicity, its topological nature in the absence of matter and the richness of its spectrum made it a handy toy model to address fundamental questions, such as the nature of horizon micro-states. The most famous example certainly lies in the observation by Brown and Henneaux that the phase space of three-dimensional gravity with a negative cosmological constant and suitable boundary conditions admits an action of the two-dimensional conformal group \cite{Brown:1986nw}, thereby suggesting a quantum description in terms of a two-dimensional CFT. This idea was made more precise when it was shown that the Bekenstein--Hawking entropy of the BTZ black hole solutions \cite{Banados:1992wn, Banados:1992gq} belonging to the phase space could be reproduced by a counting of states in a two-dimensional CFT \cite{Strominger:1997eq}. The implications of this result has been pushed in various directions over the years, attempting at identifying the precise dual field theory \cite{Witten:2007kt, Maloney:2007ud}, hinting at its non-existence in pure gravity \cite{Gaberdiel:2007ve}, and constraining the features a two-dimensional CFT dual to a three-dimensional gravity should exhibit \cite{Keller:2011xi,Hartman:2014oaa}. These results provide insightful information about how holography works in asymptotically anti-de Sitter (AdS) backgrounds, but cannot in general be translated as such to more realistic situations. For instance, the importance of the Bondi--Metzner--Sachs algebra \cite{Bondi:1962px,Sachs:1962wk,Sachs:1962zza} -- the asymptotic symmetries of four-dimensional flat space -- has been pointed out in various contexts, such as the memory effect and soft theorems (see \cite{Strominger:2017zoo} for a review of the ``infrared triangle''), and the information paradox \cite{Hawking:2016msc}, but its structure is much more involved than the conformal algebra (non-integrable charges, field dependent central extensions, Lie algebroid structure (see \cite{Barnich:2017ubf} and references therein).


In $2+1$ dimensions, non-conformal infinite-dimensional symmetry algebras have also appeared both in the asymptotic and near-horizon regions of bulk gravity theories, with \cite{Compere:2013bya, Troessaert:2013fma, Avery:201three-dimensionalja, Grumiller:2016pq, Perez:2016vqo, Donnay:2015abr,  Afshar:2015wjm, Afshar:2016wfy,  Anninos:2011vd} or without \cite{Barnich:2006av, Detournay:2016sfv, Grumiller:2017sjh} a cosmological constant. In the latter case, the asymptotic symmetries are the lower-dimensional counterpart of the Bondi-Metzner-Sachs algebra, denoted BMS$_3$. A hinderance to make use of the simplicity of pure three-dimensional gravity in the asymptotically flat case is the absence of black hole solutions \cite{Ida:2000jh}. However, cosmological solutions with non-trivial thermodynamics exist \cite{Cornalba:2002fi} and display features pointing at the relevance of an underlying BMS$_3$ symmetry \cite{Barnich:2012xq,Bagchi:2012xr}. Furthermore, modifying pure gravity by including higher-curvature terms and/or matter widens the spectrum of possible solutions 
to include certain classes of non-asymptotically AdS black objects. These include for instance the three-dimensional black string solution of Horne and Horowitz \cite{Horne:1991gn} and, more recently, an asymptotically flat hairy black hole of New Massive Gravity \cite{Oliva:2009ip}. 
Now, what is lacking in these cases is a clear definition of the corresponding phase space, its symmetries, and whether these are relevant to understand thermal properties of those objects (see however \cite{Fareghbal:2014kfa} for the role of BMS$_3$ in explaining the hairy black hole entropy, and also \cite{Carlip:2017xne} where BMS$_3$ symmetries at the horizons have been shown to capture black hole entropy).

In this note, we set the stage for such a study for a three-parameter family of black string solutions generalizing the Horne--Horowitz black string \cite{Horne:1991gn} and Witten's black hole \cite{Witten:1991yr}. It is worth pointing out that these solutions are obtained as marginal deformations of a Wess--Zumino--Witten model and as such represent exact string backgrounds \cite{Detournay:2005fz}. We start be reviewing the solution and describe some of its properties. Then, we present a consistent set of boundary conditions including these solutions and determine its asymptotic symmetry algebra of charges, the detailed derivation of which is relegated to Apps. \ref{appEOM} and \ref{appconstraintcharges}. We show in particular that these boundary conditions can be put on-shell to derive new exact time-dependent solutions to the equations of motion.
We give a CFT interpretation of these new solutions in terms of a marginal deformation of Witten's two-dimensional black hole seed. In the following section, we study thermodynamic aspects of the zero-mode black string solution, and then conclude.


\section{The charged black string configuration  }\label{backco}

The black string geometries we will be considering are the ones described in \cite{Detournay:2005fz}, consisting of a generalization of the Horne--Horowitz black string \cite{Horne:1991gn}. The latter can be viewed as the target space of a $\frac{SL (2, \mathbb{R}) \times   \mathbb{R}}{\mathbb{R}}$ gauged WZW model, or equivalently as an exact marginal deformation of the  $SL (2, \mathbb{R})$ WZW model, driven by a left-right $SL(2, \mathbb{R})$ current bilinear. The black string of \cite{Detournay:2005fz} is a generalization obtained with an extra deformation using another available exact current bilinear. As such, these geometries describe exact string theory  models

The background fields are a three-dimensional metric, a Kalb--Ramond two-form $B$, an Abelian electromagnetic gauge potential  $A$, and a dilaton $\Phi$. The dynamics of these fields is captured by the action \cite{Detournay:2005fz}
\begin{equation}
\label{lagrangian}
I=\frac{1}{16\pi G_3} \int \text{d}^3x\sqrt{-g} \left( R -4 \nabla_\mu \Phi \nabla^\mu \Phi -\frac1{12}H^2\text{e}^{-8\Phi }-\frac{k_g}8 F^2 \text{e}^{-4\Phi }+\frac{\delta c}{3\alpha'} \text{e}^{4\Phi }\right ),
\end{equation}
where $F_{\mu\nu}$ and $H_{\mu\nu\rho}$ are the components of 
\begin{eqnarray}
&F=\extd A,&\\
&H=\extd B - \frac{k_g}4 A \wedge F,&
\label{defH}
\end{eqnarray}
and  $k_g$ is the gauge coupling constant. We set Newton's constant $G_3$ to 1 throughout.

The action \eqref{lagrangian} is the low-energy effective action for the corresponding massless string degrees of freedom, written in the Einstein frame. Its extrema provide solutions generically valid as long as their length scale $L$ is much larger than $\sqrt{\alpha'}$. In some instances, as those we will be studying here, the solutions are \emph{exact, \emph{i.e.}} valid to all orders in $\alpha'$, irrespective of $L$, possibly after some finite renormalizations. Hence, $\delta c=c-d= c-3$ with $c$ the central charge of the underlying conformal sigma-model. Celebrated examples include $\text{AdS}_3$ with vanishing gauge field and dilaton, but non-zero three-form. In this case, $L$ is the anti-de Sitter radius, and $\delta c=\nicefrac{12\alpha'}{L^2}=\nicefrac{6}{k-2}$,
where $k>0$ is the level of the affine $SL (2, \mathbb{R})_k$ algebra in the corresponding sigma model. In the following
we will consider  $\delta c>0$ and parameterize it as 
\begin{equation}
\label{deltac}
\frac{\delta c}{\alpha'}=\frac{12}{L^2}.
\end{equation}
Whenever an underlying affine algebra exists, its  level $k$ will be related to $c$ as 
$\delta c\approx \frac{6}{k}$.

As already mentioned, the charged black string of \cite{Detournay:2005fz} is an exact background, reached by a double marginal deformation of the $SL (2, \mathbb{R})$ sigma model. In the Bondi gauge, its background fields read (see App. \ref{AppBS} for details): 
\begin{subequations}\label{BS4}
	\begin{align} \label{BS4met}
	\extd s^2&=4\left(- \frac{ (r-r_-) (r-r_+)}{\zeta^2} +\omega^2\right)\extd u^2-2 \frac{r}{\zeta^2} \,\extd u\,\extd r+4 \,\omega
	\,r \,\extd u\,\extd\phi+r^2\extd\phi^2
	\\  \label{matterBS}
	\Phi & =-\frac12\log \left( \frac rL\right) +\frac12 \log\zeta\\ \label{BS4A}
	A & =  \frac{4 L}{ \sqrt{k_g} \,r} \sqrt{- \omega^2 \,\zeta ^2+ r_- r_+}\,\extd u := \frac{2\alpha}{r}\extd u\\\label{BS4B}
	B & =  \frac{2 L^2 \omega\,\zeta^2}{ r}\,\extd u\wedge \extd\phi \,,
	\end{align}
\end{subequations}
where $u$ is the retarded time, $r$ the radial coordinate and $\phi$ the coordinate along the black string. For convenience, we take $\phi$ $2\pi$-periodic\footnote{Notice for later use that the coordinate $\phi$ may be non-compact. In that case, the black string charges and entropy have to be taken per unit length, because they would be divergent otherwise \cite{Kaloper:1998vw}. Also, the $\phi$ coordinate becomes the timelike in certain regions of spacetimes (inside the inner horizon), and therefore strictly speaking cannot be compactified \cite{Horne:1991gn}.}.  
In these expressions, $r_{\pm}$, $ \omega$ and $\zeta$ are arbitrary parameters, subject to conditions ensuring reality of the fields.
It should be quoted that \eqref{BS4} are formally valid for large $k$ and are thus subject to finite $\nicefrac{1}{k}$ corrections.
The solution under consideration exhibits a genuine timelike singularity at $r=0$, hidden behind two horizons located at $r=r_\pm$. 

The geometry of the above metric has been studied in \cite{Detournay:2005fz} where the full Penrose diagram was obtained.
Since we will be interested in determining asymptotic boundary conditions containing the above family of metrics, we focus for a second on what we will call the asymptotic region of the above spacetimes. Firstly, the Ricci scalar is given by
\begin{equation}
  R =-\frac{8 \zeta ^2}{r^2}-\frac{4 \zeta ^2( r_++r_-)}{r^3}+\frac{2 \zeta^2 ( \zeta ^2 \omega ^2+4  r_+ r_-)}{r^4}\,.
\end{equation}
It thus behaves like $R = O(1/r^2)$ as $r \rightarrow \infty$. Secondly, the asymptotic structure shown in the Penrose diagram \cite{Detournay:2005fz} is reminiscent of three-dimensional flat space, but is however slightly different. For the sake of the argument, let us focus on the black string metric in the original coordinates \eqref{BSmetric1} for $r_- = 0 = \omega$, $\zeta = 1$. Defining $u = t/L - r^* = \tan U$, $v = t/L + r^* = \tan V$, with tortoise coordinate $r^* = 1/4 \ln(4 |r-r_+|)$, the large radius behaviour of the metric is given by
\begin{equation}
 \text{d}s^2 = \frac{4 \, r^2 (U,V)}{\cos^2 U \cos^2 V} \left(- \extd U \extd V + \cos^2 U \cos^2 V \frac{\extd x^2}{L^2}\right)
\end{equation}
with $|U| \leq \frac{\pi}{2}$ and $|V| \leq \frac{\pi}{2}$. One could then define different regions in the Penrose diagram of the Schwarzschild patch of the black string analogous to those of three-dimensional Minkowski space: $i_0 \equiv \{U = -\frac{\pi}{2}, V = \frac{\pi}{2}\}$, $i^\pm \equiv \{U = V=  \pm \frac{\pi}{2}\}$, ${\mathcal I}^+ \equiv \{V =  \frac{\pi}{2}, |U| <  \frac{\pi}{2} \}$, ${\mathcal I }^- \equiv \{U =  -\frac{\pi}{2}, |V| <  \frac{\pi}{2} \}$, with the difference that the $xx$ component of the unphysical metric is $\cos^2 U \cos^2 V$ instead of $\sin^2 (V-U)$ in the Minkowski case. The metric is thus not asymptotically flat in the usual sense. Thirdly\footnote{Ph. Spindel, private communication.}, the study of the geodesics in the background (\ref{BSmetric1}) shows that spacelike geodesics end at $i_0$ (for $r\rightarrow \infty$, with $t$ and $x$ finite), while null geodesics reach $r \rightarrow \infty$ for infinite values of $t$ and $x$ corresponding to ${\mathcal I}^\pm$.

\section{Phase space}
In this section, boundary conditions including the black string, are presented and shown to be consistent. 
The two first paragraphs stress the key points of the reasoning leading to our set of boundary conditions. The interested reader is encouraged to read the details in App. \ref{appEOM} and \ref{appconstraintcharges}.  
The boundary conditions are displayed in Sec. \ref{SSectPhasespace}. Their consistency, asymptotic symmetry algebra and associated conserved charges are worked out in Sec.~\ref{Charge algebra}.


\subsection{Bondi gauge and equations of motion}	
In order to gain some insight into the type of boundary conditions that could be imposed with the action (\ref{lagrangian}) at hand, we follow and generalise the strategy outlined in App. A of \cite{Barnich:2015jua}, originally inspired by \cite{Winicour85}, 
in order to (partially) solve the equations of motion asymptotically. The details are presented in App. \ref{appEOM}. We use coordinates $(u,r,\phi)$ and the gauge fixing ansatz
\begin{align}\label{ansa}
	g_{\mu\nu} &= \left(
	\begin{array}{ccc}
		r^2 U^2+L^2 \text{e}^{\beta} V& -L \, \text{e}^{\beta} & r^2\,U \\
		-L \, \text{e}^{\beta} & 0& 0 \\
		r^2\,U & 0 & r^2
	\end{array}
	\right) \,, \qquad 
	A_\mu=\left(A_u \  A_\phi \right)
\end{align}
with  $U$, $V$, $\beta$, $A_u$, $A_\phi$ functions of $(u,r,\phi)$. 

First of all, in three dimensions, the equation of motion for the three-form, 
\beq
\nabla_\mu
H^{\mu\nu\rho}-8H^{\mu\nu\rho} \nabla_\mu \Phi =0
\eeq
 is automatically solved for 
\begin{equation}
H=\Omega \text{ e}^{8\Phi}\sqrt{-g}  
\text{d}u\wedge
\text{d}r\wedge
\text{d}\phi\,
\label{genH}
\end{equation}
where $\Omega$ is a constant. 

The equations of motion for the metric, the electromagnetic field and the dilaton are
\begin{align}
	&E_{\mu \nu} := G_{\mu \nu} - T_{\mu \nu} = 0 \\ \nonumber
	& \mbox{with} \; T_{\mu\nu}= 
	4\left(\nabla_\mu\F\nabla_\nu \F -\frac12g_{\mu\nu}(\nabla_\mu\F)^2\right)+
	\frac14 \text{e}^{-8\F}\left (H_{\mu ab}H_\nu^{{\phantom{\nu}}ab}-\frac16g_{\mu\nu}H^2\right )\\
	&\qquad \qquad +
	\frac{k_g}4 \text{e}^{-4\F}\left(F_{\mu a}F_\nu^{\phantom{\nu}a}-\frac14 g_{\mu\nu} F^2\right )
	+
	\frac {2}{L^2} \text{e}^{4\F}\, \\
	&  {\cal J}^{\nu} := \nabla_\mu {\cal F}^{\mu \nu} :=  \nabla_\mu [\text{e}^{-4 \Phi} F^{\mu \nu} - \Omega \, \eta^{\mu \nu \rho} A_\rho] = 0 \\
	& E_\Phi := 8 \nabla^\mu\nabla_\mu \Phi + \frac{1}{2} k_g F^2 \text{e}^{-4 \Phi} + \frac {16}{L^2}\, \text{e}^{4 \Phi} - 4 \Omega^2 \text{e}^{8 \Phi} = 0\,.
\end{align}
The upshot of \cite{Winicour85,Barnich:2015jua} is that one can determine the asymptotic behaviors of the fields given an ansatz for $A_\phi$ and $\Phi$ using the so-called main equations of motion (see App. \ref{appEOM}). Then, we partially solve other equations of motion (until we reach equations involving either sub-leading components or non linear PDEs). 
Writing
\begin{equation}
 \label{ansadil}
	\Phi = -\frac{1}{2} \log \frac{r}{L} + f
\end{equation}
with $f$ a function of $(u,r,\phi)$, we take the ansatz to be 
\begin{align}
	A_\phi(u,r,\phi)&= a_{00} (u,\phi)+ a_{11}  (u,\phi)\frac {\log \left( \frac rL \right) }r + a_{01} (u,\phi) \frac1r +O\left( \frac {\log^2\left( \frac rL \right)} {r^2}\right) \\ 
	 f (u,r,\phi)&= f_{00} (u,\phi)+ f_{11}  (u,\phi)\frac {\log \left( \frac rL \right) }r + f_{01} (u,\phi) \frac1r +O\left( \frac {\log^2\left( \frac rL \right)} {r^2}\right) \,.
\end{align}
The equations of motion then unambiguously fix the remaining functions:
\begin{subequations}\label{FOEOM}
\begin{align}  \label{EOMbeta}
		\beta =& \log \left( \frac rL \right) + \beta_{00} + \beta_{11}\frac {\log \left( \frac rL \right) }r  + \beta_{01} \frac{1}{r} + O\left(\frac{\log^2 \left( \frac rL \right) }{r^2}\right)\\ \nonumber
&\text{with } \beta_{00}=-4f_{00}+b_0\,, \quad  \beta_{01} = - 4 f_{01}, \quad \beta_{11} = - 4 f_{11}\\ \nonumber
& \\ \label{EOMU}
U = &U_{00} + U_{21} \frac{\log^2 \left( \frac rL \right) }{r} + U_{11} \frac{\log \left( \frac rL \right) }{r}  + U_{01} \frac{1}{r} + O\left(\frac{\log^3 \left( \frac rL \right) }{r^2} \right)\\  \nonumber
&\text{with } \begin{aligned}[t] &U_{00}  =U_0 \text{e}^{-2 f_{00}}+2 \text{e}^{\beta_{00}}\p_\phi f_{00}+\text{e}^{-2f_{00}} \p_u F\,, \text{where } F=\int_0^\phi \text{e}^{2f_{00}(u,\theta)}\extd\theta \,,\\
& U_{21}=	-2\text{e}^{\beta_{00}}\left( 2f_{11} \p_\phi f_{00} - \p_\phi f_{11}\right)  \,, \\
&U_{11} =-4 \text{e}^{\beta_{00}} \left( 2f_{01} \p_\phi f_{00} - \p_\phi f_{01}\right) \,,
\end{aligned}
\nonumber
& \\ \label{EOMAu}
A_u =&  \alpha_{00} +  \alpha_{21} \frac{\log^2 \left( \frac rL \right) }{r} + \alpha_{11} \frac{\log \left( \frac rL \right) }{r}  + \alpha_{01} \frac{1}{r} + O\left(\frac{\log^3 \left( \frac rL \right)}{r^2}\right) \\ \nonumber
&\text{with } 	
\begin{aligned}[t] & \p_\phi \alpha_{00}=\p_u a_{00}\,,\\
 & \alpha_{21}= \p_\phi \Big[ \frac12 \text{e}^{-4f_{00}+b_0} a_{11} \Big] 	\,, \\ 
 & \alpha_{11}=\p_\phi \Big[  \text{e}^{-4f_{00}+b_0} a_{01} \Big]+ a_{11} U_{00}  \,,
 \end{aligned}
 \end{align}
\begin{align}
\label{EOMV}
V = & \bar V_{01}\, r  + V_{30} \log^3 \left( \frac rL \right)  + V_{20}\log^2 \left( \frac rL \right)+ V_{10}\log \left( \frac rL \right)+ V_{00}   +   O\left(\frac{\log^4 \left( \frac rL \right) }{r}\right)\\ \nonumber
&\text{with } 
\begin{aligned}[t] & \bar V_{01} = \frac2L \partial_\phi U_{00} -\frac{4}{L} \text{e}^{b_0} (1- 3 \text{e}^{-4f_{00}} (\partial_\phi f_{00})^2 +\text{e}^{-4f_{00}} \p_\phi^2 f_{00} )\\
& V_{30}=\frac{2}{3L} \text{e}^{-4f_{00}+b_0} \Big(-6 \p_\phi f_{00} \p_\phi f_{11}+\left(8 (\p_\phi f_{00})^2-2 \p_\phi^2 f_{00}\right)  f_{11}+\p_\phi^2 f_{11}\Big)\\ 
& V_{20}=	\frac2L \text{e}^{-4f_{00}+b_0} \Big[ -6\p_\phi f_{00}\p_\phi f_{01}+
		f_{11} ( -4(\p_\phi f_{00})^2 +2 \p_\phi^2f_{00}  ) \\ 
& \qquad \qquad \qquad \qquad \qquad +\p_\phi^2 f_{01}
		+f_{01}(8(\p_\phi f_{00})^2-2\p^2_\phi  f_{00}	)  \Big]\\ 
& V_{10}= \frac{1}L  \Big[ -8 \text{e}^{-4f_{00}+b_0} f_{01}\left(2 (\p_\phi f_{00})^2-\p_\phi^2 f_{00}\right) +\p_\phi U_{01} \Big],
\end{aligned}
	\end{align}
\end{subequations}
where $b_0$ and $U_0$ are functions of $u$ while all other functions depend on $(u,\phi)$.

The $H$-field is known, but it will be useful for the computation of the charges to derive the asymptotic form of the $B$-field \eqref{defH}.
Choosing the gauge for which $B_{ur}=B_{r\phi}=0$ and taking the arbitrary term of order $1$ to be zero, it is
\begin{align}\nonumber
	B_{u\phi}(u,r,\phi)=&
	\frac{1}{8} k_g a_{00}  \p_\phi \left(\text{e}^{-4f_{00}+b_0} a_{11}
	\right) 
	 \frac {\log^2 \left( \frac rL \right) }r \\ \nonumber
	 &+ 
	\frac{1}{4} k_g \left(-a_{11 } \alpha_{00} +a_{00} \left( a_{11}U_{00}+\p_\phi (  \text{e}^{-4f_{00}+b_0}a_{01}) \right) \right) 
	\frac {\log \left( \frac rL \right) }r\\
	& + \frac{1}{4} \left(k_g a_{00} \alpha_{01}-k_g  a_{01} \alpha _{00}+4L^4 \Omega \,  \text{e}^{4f_{00}+b_0}\right)\frac1r  + O\left( \frac {\log^{3}\left( \frac rL \right)} {r^2}\right) \,.
\end{align}


\subsection{No BMS$_3$}
We will ultimately be interested in determining a consistent set of boundary conditions and its symmetries. 
Given any field configuration $\Psi=(g,\Phi,A,B)$ included in the boundary conditions, asymptotic symmetries are the transformations $\Xi $ such that $\Psi +\delta_\Xi \Psi$ remains included in the boundary conditions and leading to well-defined charges. More precisely, we have that $\delta_\Xi \Psi=(\mathsterling_\xi g,\mathsterling_\xi\Phi,\mathsterling_\xi A +\extd \lambda,\mathsterling_\xi  B+\extd \Lambda -\frac{k_g}4 \extd \lambda\wedge A )$ with $\xi$
a vector field, $\lambda$ a scalar and $\Lambda$ a one-form. 

Let us give a first look at the allowed symmetries. By first imposing to preserve the Bondi gauge (from \eqref{ansa}), we get
\begin{align}\label{key}
	&\mathsterling_\xi g _{rr} =0 \Rightarrow \xi^u= \xi^u(u,\phi)\\
	&\mathsterling_\xi g _{\phi\phi}=0  \Rightarrow \xi^r=-r\,U\,  \p_\phi \xi^u-r\, \p_\phi \xi^\phi\\
	&\mathsterling_\xi g _{r\phi}=0  \Rightarrow \xi^\phi= \p_\phi \xi^u\int^r \frac{\text{e}^{\beta}}{\rho^2}\extd\rho+Y(u,\phi)\,. 
\end{align}
Using the results \eqref{FOEOM}, we have 
\begin{align} \nonumber
	\xi^\phi = & \text{e}^{-4f_{00}+b_0}  \p_\phi \xi^u \log \left( \frac rL \right) +Y(u,\phi)
+ 4 \text{e}^{-4f_{00}+b_0}  f_{11}  \p_\phi \xi^u\frac {\log \left( \frac rL \right) }r  \\
&+ 4  \text{e}^{-4f_{00}+b_0} (f_{01}+f_{11}) \p_\phi \xi^u \frac{1}{r} + O\left(\frac{\log^2 \left( \frac rL \right) }{r}\right)\,.
	\end{align}
Then, we consider the dilaton transformation 
\begin{align} \nonumber
\mathsterling_\xi \Phi = & 
\left(\frac12 U-r\,U\,\p_r f\right) \p_\phi \xi^u +\p_u \, \xi^u 
+\left(\frac12-r\p_rf\right)\p_\phi \xi^\phi+\p_\phi f\, \xi^\phi 
\,.
	\end{align}

Using again \eqref{FOEOM}, one finds that the leading term of this transformation is of order $\log(r/L)$. This term spoils the ansatz \eqref{ansadil} and must vanish, which implies that 
\begin{equation}\label{keytext}
	\xi^u=\int^\phi  c_1(u) \text{e}^{2 f_{00}(u,\Theta)} \, \extd \Theta+X(u)\,.
\end{equation}
The periodicity of the $\phi$ variable forces us to take $c_1=0$. 
The vector field thus takes the form 
\begin{equation}\label{KVBC}
	\xi=X(u)\p_u - r \, \p_\phi Y(u,\phi) \p_r+ Y(u,\phi) \p_\phi \,. 
\end{equation}
The preservation of the dilaton ansatz implies that the supertranslation generator of the BMS$_3$ algebra is not included in the asymptotic symmetries (since this would require a $\phi$-dependence in $\xi^u$, see \emph{e.g.} Eq. (6) of \cite{Bagchi:2012yk}). It would be interesting to relax the periodicity condition on $\phi$ and to allow for a more general ansatz for $\Phi$ \eqref{ansadil}, possibly including non-integer powers of $r$.


It is easily checked that the vector field \eqref{KVBC} preserves the orders of the three other components of the metric and the dilaton. 
We now consider the transformation of the Maxwell field. Preserving its form implies
\begin{equation}\label{lamMaxBC}
   \partial_r \lambda = 0.
\end{equation}
Finally, the condition on the $B$-field leads to 
\begin{align}\label{LambdaEOM}
	& \Lambda=\extd M + \Lambda_\phi\extd \phi
\end{align} 
with $\Lambda_\phi$ being constant and $M$ an arbitrary function. 


After having solved partially the equations of motion and put restrictions on the potential asymptotic symmetries, the next step is to compute the charges and impose them to be finite and integrable. This has restricted the boundary conditions and has eventually led to a consistent of set of boundary conditions. For clarity reasons, we have decided to present in main text the obtained phase space and compute the charges for this set. We relegate in App. \ref{appconstraintcharges} this step of the construction of the boundary conditions.

\subsection{Phase space and symmetries}\label{SSectPhasespace}
In this section we present a consistent set of boundary conditions including the black string solutions and their symmetries. We spell out in  App. \ref{appconstraintcharges} the details that led us to this particular set.

 
The boundary conditions are (all coordinate-dependences are explicit):
\begin{subequations}\label{BC}
	\begin{align}
	A_\phi(u,r,\phi)&= a_{11}(u,\phi) \frac{\log \left( \frac rL \right)}{r}+ a_{01}(u,\phi) \frac1r +O\left( \frac {\log^2\left( \frac rL \right)} {r^2}\right) \\
	f (u,r,\phi)&= f_{00} (\phi) +O\left( \frac {\log^2\left( \frac rL \right)} {r^2}\right) \\
	\nonumber&\\
	\beta(u,r,\phi)&= \log \left( \frac rL \right)  -4 f_{00}(\phi) +O\left ( \frac{\log^2 \left( \frac rL \right) }{r^2} \right )\\
	U (u,r,\phi)&=  2 \text{e}^{-4 f_{00}(\phi)} \p_\phi f_{00}(\phi) +   u_{01} \frac{1}{r} + O\left (\frac{\log^3 \left( \frac rL \right) }{r^2}\right)\\ \nonumber
		\nonumber&\\
	A_u(u,r,\phi) &=   \alpha_{21}(u,\phi) \frac{\log^2 \left( \frac rL \right) }{r} + \alpha_{11}(u,\phi) \frac{\log \left( \frac rL \right) }{r} + \alpha_{01}(u,\phi) \frac{1}{r} + O\left (\frac{\log^3 \left( \frac rL \right)}{r^2}\right)  \\ \nonumber
	&\text{with } 	
	\begin{aligned}[t] 
	& \alpha_{21}(u,\phi)= \p_\phi \left[ \frac12 \text{e}^{-4f_{00}(\phi)}a_{11}(u,\phi) \right]  	\,, \\ 
	& \alpha_{11}(u,\phi)= \p_\phi \left[ \text{e}^{-4f_{00}(\phi)} a_{01}(u,\phi) \right]+ 2a_{11}(u,\phi)\text{e}^{-4f_{00}(\phi) }\p_\phi f_{00}(\phi)   \,,
	\end{aligned}
	\\ \nonumber
	& \\ 
	V(u,r,\phi)& =- \frac{4}L \left( 1 + \text{e}^{-4 f_{00}(\phi)} (\p_\phi f_{00}(\phi))^2\right) \, r  + V_{00}(u,\phi)   +   O\left( \frac{\log^4 \left( \frac rL \right) }{r}\right) \,.
	\end{align}
\end{subequations}
The $B$-field takes the form 
\begin{equation}\label{key}
B_{u\phi}(u,\phi) =\text{e}^{4f_{00}(\phi) }  \Omega L^4 \frac1r+ O\left( \frac {\log^{3}\left( \frac rL \right)} {r^2}\right) \,.
\end{equation}
Translating in terms of metric components, we have the following falloffs 
\begin{align}
& g_{uu}= - 4 \text{e}^{-4f_{00} (\phi)} r^2 + \text{e}^{-4f_{00} (\phi)}(L\, V_{00}(u,\phi) +4 u_{01} \p_\phi f_{00}(\phi )) r +O\left(\log^4\left(\frac rL\right)\right)\\
& g_{ur}=-\text{e}^{-4f_{00} (\phi)} r + O\left( \frac{\log^2 \left( \frac rL \right) }{r}\right)\\
& g_{u\phi}=2 \text{e}^{-4f_{00}(\phi)} \p_\phi f_{00}(\phi)r^2+u_{01} r+ O\left(\log^3\left(\frac rL\right)\right) \,. 
\end{align}
The black string \eqref{BS4} is included in the phase space with 
\begin{align}\nonumber
f_{00}&=\frac12 \log \zeta\,, \quad V_{00}= \frac4L(r_++r_-) \,, \quad  u_{01}=2\omega \,, \\ \alpha_{01}&=  \frac{4 L}{ \sqrt{k_g} } \sqrt{- \omega^2 \,\zeta ^2+ r_- r_+}\,, \quad \Omega=\frac{2\omega}{L^2} 
\end{align}
and the other functions and subleadings to zero.

Notice that these boundary conditions are presented in a particular gauge, and by having partially solved the equations of motion. Transposed in the pure AdS$_3$ context, this would be somewhere in between the Brown--Henneaux boundary conditions of \cite{Brown:1986nw} and the completely gauged-fixed, on-shell solutions of Ba\~nados \cite{Banados:1998gg} though, since the theory at hand presently does exhibit propagating degrees of freedom, we do not expect to be able to write down the most general exact solution incarnating our boundary conditions.

The asymptotic symmetry parameters preserving these boundary conditions are in the form of a triplet $(\xi, \lambda, \Lambda)$ with
\begin{equation}\label{key}
\xi= X \p_u- r \p_\phi Y(\phi )\p_r+Y(\phi)\p_\phi\,, \quad \lambda \in \mathbb{R} \,,\quad \Lambda=\Lambda_\phi \, \extd \phi \,. 
\end{equation}
For future computations, it is relevant to write the transformations $\delta$ of some fields under the action of the asymptotic symmetry generator $a=(\xi,\lambda,\Lambda)$, 
\begin{subequations}\label{deltaa}
	\begin{align}
	& \delta_a f_{00}= \frac{\p_\phi Y}2+Y\,\p_\phi f_{00} \,,\quad  \delta_a \Omega=0\,, \quad \delta_a u_{01}=0 \\ \label{deltaV00}
	& \delta_a V_{00}=V_{00} \p_\phi Y + \p_\phi V_{00} Y + \p_u V_{00} X -2\frac{u_{01}}L \p_\phi^2 Y \\
	&  \delta_a a_{01}=2a_{01} \p_\phi Y -a_{11}\p_\phi Y+ \p_\phi a_{01} Y + \p_u a_{01} X\\
	& \delta_a a_{11}=2a_{11} \p_\phi Y + \p_\phi a_{11} Y + \p_u a_{11} X\\ 
	& \delta_a \alpha_{01}=\alpha_{01}\p_\phi Y+ \p_\phi \alpha_{01} Y + \p_u \alpha_{01} X- 2\text{e}^{-4f_{00}} a_{11} \p_\phi f_{00} \p_\phi Y -\p_\phi (a_{01}\text{e}^{-4f_{00}} ) \p_\phi Y \,. 
	\end{align}
\end{subequations}

In the following section, we show that the charges associated with these parameters are finite and integrable on the above phase space. Moreover, we will make a small restriction of the phase space to obtain conserved charges. Finally, we will determine the asymptotic symmetry algebra.


\subsection{Charge algebra}
\label{Charge algebra}
\paragraph{General expressions of the charges}
The explicit expressions of the charges for the theory \eqref{lagrangian} are partially known. 
The contributions from the gravitational part \cite{Iyer:1994ys, Abbott:1982jh,Anderson:1996sc}, the Maxwell part \cite{Barnich:2005kq}, the two-form and dilaton parts \cite{Compere:2007vx, Detournay:201two-dimensionalz} are displayed in the literature. The extra contribution comes from the piece proportional to $k_g$ in \eqref{defH}. To keep track of it, we parametrize the factor $k_g/4$ by $k$ in the definition \eqref{defH}, $H=\extd B-k A\wedge F$. To determine it, we use the BBC method \cite{Barnich:2001jy, Barnich:2007bf} (see also App. A of \cite{Anninos:2011vd} and \cite{Chen:2013aza} for a brief summary, and \cite{Compere:2018aar} for a pedagagocial account).

The first ingredient is to consider the weakly vanishing N\oe ther current, denoted $S$. 
It is defined as the Lagrangian variation with respect to all fields of the theory times their reducibility parameters $R$:
\begin{align}
	S= \frac{\delta L}{\delta g}R_{g} +\frac{\delta L}{\delta A}R_{A}+\frac{\delta L}{\delta B}R_{B}+\frac{\delta L}{\delta \F}R_{\F}=  \left(S_{EH}+S_{T_{\mu\nu}}\right) +S_{A}+ S_{B}+S_{\F} \,.
\end{align}
For the theory \eqref{lagrangian}, we get 
\begin{subequations}
	\begin{align}
		& S^\mu_{T_{\mu\nu}}= \frac{\sg}{8\pi}T^{\mu\nu}_{\text{matter}}\,\xi_\nu \,\quad \text{ with}\\ \nonumber
		&\quad  T^{\text{matter}}_{\mu\nu}= 
		4\left(\nabla_\mu\F\nabla_\nu \F -\frac12g_{\mu\nu}(\nabla_\mu\F)^2\right)+
		\frac14 \text{e}^{-8\F}\left (H_{\mu ab}H_\nu^{{\phantom{\nu}}ab}-\frac16g_{\mu\nu}H^2\right )\\
		&\quad \qquad \qquad +
		\frac{k_g}4 \text{e}^{-4\F}\left(F_{\mu a}F_\nu^{\phantom{\nu}a}-\frac14 g_{\mu\nu} F^2\right )
		+\frac2{L^2} \text{e}^{4\F}g_{\mu\nu}\\ \nonumber
		& S^\mu_A= 
		\frac{\sg}{16\pi} \frac{k_g}2\left ( \nabla_\sigma F^{\sigma\mu}-4 (\nabla_\sigma \F)F^{\sigma\mu} \right ) \text{e}^{-4\F}\bar\lambda\\
		&\qquad 
		+ k \frac{\sg}{16\pi} \left (
		H^{\tau\mu\rho} F_{\rho\tau} + A_\tau \left(\nabla_\rho H^{\tau\mu\rho}-8 \nabla_\rho \F H^{\tau\mu\rho} \right)  \right )\text{e}^{-8\F} \bar\lambda\\
		& S^\mu_B=  \frac{\sg}{16\pi}
		(\nabla_\tau H^{\tau\mu\sigma} -8 (\nabla_\tau \F) H^{\tau\mu\sigma}  )\text{e}^{-8\F}
		\bar \Lambda_\sigma\\
		& S^\mu_\F=0,
	\end{align}
\end{subequations}
where $\bar\lambda=A_\rho\xi^\rho+\lambda$ and $\bar \Lambda_\sigma=B_{\rho\sigma}\xi^\rho+\Lambda_\sigma-k \lambda\,A_\sigma$. 
Then, we apply a contracting homotopy operator to this weakly vanishing N\oe ther current to obtain a one-form potential  $k^{[\mu\nu]}$ (in 3 dimensions). For a second order theory, this operation can be written in the following way: 
\begin{equation}
	k_{\xi,\lambda,\Lambda}^{[\mu\nu]}=
	\frac12 \delta \varphi^i \frac{\partial}{\partial \varphi^i_{,\nu}} S_{\xi,\lambda,\Lambda}^\mu +
	\left( \frac23 \partial_\lambda \delta\varphi^i-\frac13\delta \varphi^i \partial_\lambda  \right ) \frac{\partial}{\partial \varphi^i_{,\lambda\nu}} S_{\xi,\lambda,\Lambda}^\mu -(\mu\leftrightarrow \nu)\, ,
\end{equation}
where $\varphi$ are the fields of the theory labelled by the index $i$ and $\delta \varphi$ their variation. 

For the present work, it is useful to introduce the following tensor
\begin{equation}
	\lozenge^{\alpha\beta\gamma}_{\tau\mu\rho}= 
	\delta^\alpha _\tau \delta^\beta_\mu \delta^\gamma_\rho  +
	\delta^\alpha _\mu \delta^\beta_\rho \delta^\gamma_ \tau +
	\delta^\alpha _\rho \delta^\beta_ \tau\delta^\gamma_ \mu \,, \quad \lozenge^{\tau\mu\rho\,\nu\sigma\lambda} = \lozenge_{\alpha\beta\gamma}^{\tau\mu\rho} g^{\alpha\nu}g^{\beta\sigma}g^{\gamma\lambda}\,. 
\end{equation}
The contributions of the matter fields $\Phi,A,B$ to the energy-momentum tensor are (with $a_\mu := \delta A_\mu$,  $b_{\mu \nu} := \delta B_{\mu \nu}$)
\begin{align}\label{kT}
	k^{T_{\mu\nu}}= &8 \p^\mu \Phi \, \xi^\nu\,f+ (g^{\mu\tau}\xi_\sigma H^{\sigma\nu\rho})(\frac12 b_{\rho\tau}-k\,a_{[\rho}A_{\tau]})\text{e}^{-8\Phi}\nonumber \\
	& + \frac {k_g}4(-g^{\mu\rho}F^{\sigma\nu}+2F^{\mu\rho}g^{\sigma\nu}-g^{\sigma\rho}F^{\mu\nu})\xi_\sigma\,a_\rho\, \text{e}^{-4\Phi} - (\mu\leftrightarrow\nu) \,.
\end{align}
The contribution coming from Einstein--Maxwell plus dilaton theory is
\begin{align}
	k^{F,\Phi}=& \nonumber
	k_g \Big(
	\nabla^\mu\F\, a^\nu + F^{\mu\nu}\, f + \frac12 F^{\mu\gamma}h^\nu_{\gamma} -\frac18 F^{\mu\nu}h
	\Big) \bar\lambda \text{e}^{-4\F} \\
	&+k_g g^{\mu\kappa}  g^{\lambda\nu}
	\left(\frac12 \p_\lambda a_\kappa- \frac14a_\kappa\p_\lambda\right) \Big(  \bar\lambda \text{e}^{-4\F}\Big)- (\mu\leftrightarrow\nu) \,. \label{K1} 
\end{align}
In the part coming from $H^2$ in the Lagrangian, we have the pure three-form part, proportional to $\bar \Lambda$ and the contribution of the Chern-Simons coupling
\begin{align}
	k^{H}=& \nonumber
	\big(-4\lozenge^{\tau\mu\sigma\,\nu\beta\gamma} \p_\tau \Phi\,b_{\beta\gamma}+4H^{\mu\nu\sigma}\,f
	+\frac14H^{\mu\nu\sigma} h\big)\text{e}^{-8\Phi}(\bar \Lambda_\sigma-k\, \bar \lambda\,A _\sigma)\\ \nonumber
	&- \lozenge^{\sigma \mu(\nu\,\lambda)\kappa\omega}\Big(\frac23 \p_\lambda b_{\kappa\omega}- \frac13b_{\kappa\omega}\p_\lambda\Big) 
	\Big(
	\text{e}^{-8\Phi}(\bar \Lambda_\sigma -k\, \bar \lambda\,A _\sigma)\Big) \\ \nonumber
	&+ k \Big(F^{\sigma\mu}a^\nu+\frac12F^{\mu\nu}a^\sigma+\lozenge^{\sigma\mu\tau\,\alpha[\nu\kappa]}\p_\tau A_\alpha\,a_\kappa+8\lozenge^{\sigma\tau\mu\,\alpha[\nu\kappa]}\p_\tau \F A_\alpha \, a_\kappa
	\Big) \text{e}^{-8\Phi}(\bar \Lambda_\sigma-k\, \bar \lambda\,A _\sigma) \\ \nonumber
	&+k\big(\lozenge^{\sigma\mu\lambda\,\alpha[\nu\kappa]}+ \lozenge^{\sigma\mu\nu\,\alpha[\lambda \kappa]}\big) \left(\frac23 \p_\lambda a_\kappa- \frac13a_\kappa\p_\lambda\right)
	\Big(   A_\alpha \text{e}^{-8\Phi}(\bar \Lambda_\sigma-k\, \bar \lambda\,A _\sigma) \Big)\\ \nonumber
	&+k \Big( -F^{ \mu\rho}  b_{\rho}^{\,\,\,\nu}  
	+ H^{\mu\nu\tau} a_\tau 
	+2 k\,F^\mu_{\,\,\,\tau}A^{[\nu} a^{\tau]} \Big)\bar\lambda \text{e}^{-8\F}\\ \label{kH}
	&- (\mu\leftrightarrow\nu)\,.
\end{align}

Finally, the infinitesimal charge difference between a configuration $\Psi$ and $\Psi + \delta \Psi$ is obtained by integrating this one-form over a surface at infinity:
\begin{equation}\label{intdk}
	\delta H_{(\xi,\lambda,\Lambda)}=\int_{S_\infty} k_{(\xi,\lambda,\Lambda)}(\Psi,\delta \Psi)\,,
\end{equation}
with
\begin{equation}\label{kBS}
	k_{(\xi,\lambda,\Lambda)}[\delta \Psi, \Psi] = k^{G}+  k^{T_{\mu\nu}} + k^{F,\Phi} + k^{\Phi} + k^{H} \,
\end{equation}
with $k^G$ is the gravitational part \cite{Abbott:1982jh, Iyer:1994ys, Anderson:1996sc} and the other contributions are given by \eqref{kT}, \eqref{K1} and \eqref{kH}. When computing the charges in the next section, we will restore $k$ to $k_g/4$. 
The finite charge difference is then obtained by integrating along a path in configuration space.

\paragraph{Charges}

The integrated charges for the boundary conditions \eqref{BC} are given by 
\begin{subequations}\label{charges}
\begin{align} \label{chargeMax}
H_{(-rY'\p_r+Y\p_\phi,0,0)}&=\frac1{16\pi}\int_0^{2\pi} \extd\phi \,\text{e}^{4f_{00}(\phi)} u_{01} Y(\phi)\\ 	\label{chargepu}
	H_{(\p_u,0,0)}&=\frac L{16\pi}\int_0^{2\pi} \extd\phi\,  V_{00}(u,\phi)\,\quad \\	
	 H_{(0,0,\Lambda_\phi \extd\phi )}&=\frac1{16\pi }\int_0^{2\pi} \extd \phi\,  \Omega\,\Lambda_\phi\,\\
	 H_{(0,\lambda,0)} &= \frac{k_g\,\lambda }{32\pi L^2} \int_0^{2\pi} \extd \phi\, \mathcal A(u,\phi)\,  ,\label{chargellambda}
\end{align}
\end{subequations}
where 
\begin{equation}\label{key}
\mathcal A:= \alpha_{01}(u,\phi) -2 \text{e}^{-4f_{00}(\phi)} a_{01}(u,\phi)\p_\phi f_{00}(\phi)\,.   
\end{equation}


The charges \eqref{chargepu} and \eqref{chargellambda} are not conserved, as they are explicitly $u$-dependent. 
This $u$-dependence can be interpreted as gravitational and electromagnetic news, respectively. Notice that, contrary to the more familiar situations of pure gravity in four dimensions \cite{Wald:1999wa} or Einstein- Maxwell theory in three dimensions \cite{Barnich:2015jua} where once charges are made integrable they become automatically conserved, here we can make them integrable while still non-conserved (compare for instance to Eq. (3.1) of \cite{Barnich:2011mi} for the former case and to Eq. (4.4) of \cite{Barnich:2015jua} for the latter).
However, a further restriction of the phase space leads to conserved charges as we now explain. 

Using the periodicity of $\phi$, we decompose the functions $V_{00}$ and $\mathcal A$ in modes. It is sufficient to demand that their zero modes be constant to have conserved charges
\begin{equation}\label{restconscharges}
V_{00}(u,\phi)=V_0+\sum_{n\neq 0 } V_n(u)\,\text{e}^{i\,n\,\phi}\,,\qquad \mathcal A(u,\phi)=\mathcal A_0+\sum_{n\neq0} \mathcal A_n(u)\,\text{e}^{i\,n\,\phi} \,.
\end{equation}

This is only a consistent requirement if the condition is preserved by the asymptotic symmetries which is indeed the case as we now show. 
Under the asymptotic group, $V_{00}$ transforms as \eqref{deltaV00}, where (taking $Y(\phi)=\text{e}^{i\,k\phi}$),  
\begin{align} \nonumber
\delta_a V_{00}&= 
(\sum_{n} V_n \, \text{e}^{i\,n\,\phi}) i \,k\, \text{e}^{i\,k\,\phi} +( \sum_{n} V_n \,i\,n\,\text{e}^{i\,n\,\phi}) \text{e}^{i\,k\,\phi}+X\, (\sum_{n} \p_u V_n\, \text{e}^{i\,n\,\phi}) 
+2\frac{u_{01}}L\,k^2  \text{e}^{i\,k\,\phi} \, 
\\  
&= 
\sum_{m} \left(V_{m-k}\,i\,m+X\,\p_u V_m+\frac{2u_{01}} L k^2 \delta_{k-m}\right)  \text{e}^ {i\,m\,\phi} \,. 
\end{align}
Thus, 
\begin{equation}\label{key}
\delta_a V_0=0 \,, \qquad \delta_a V_{n}(u)= V_{n-k} \,i\, n+X\, \p_u V_n+\frac{2u_{01}} L k^2 \delta_{k-n} \,,\text{for $n\neq0$} \,,
\end{equation}
 and the form of the field $V_{00}$ is indeed preserved by the asymptotic symmetries. 
The variation of $\mathcal A$ is, using \eqref{deltaa},
\begin{align} 
\delta_a \mathcal A= &  \mathcal A\, \p_\phi Y + \p_\phi \mathcal A\, Y +\p_u \mathcal A \, X -\p_\phi [ a_{01} \text{e}^{-4f_{00}} \p_\phi Y].
\end{align}
Decomposing $ a_{01}\text{e}^{-4f_{00}}$ in modes the same way, one has 
\begin{align} \nonumber
\delta_a \mathcal A= &   \sum_{m} (\mathcal A_{m-k}\,i\,m+X\,\p_u \mathcal A_m)  \text{e}^ {i\,m\,\phi} +\sum_n  (a_{01}\text{e}^{-4f_{00}})_n\, k(n+k) \text{e}^{i (n+k) \phi} \\
=&\sum_{m} \left(\mathcal A_{m-k}\,i\,m+X\,\p_u \mathcal A_m + (a_{01}\text{e}^{-4f_{00}})_{m-k}\, k\, m\right)  \text{e}^ {i\,m\,\phi} \,. 
\end{align}
Therefore, the zero mode of $\mathcal A$ stays constant under the action of the asymptotic symmetries. 

The restriction \eqref{restconscharges} on the boundary conditions \eqref{BC} leads to conserved, integrable and finite charges. In the following, we compute the algebra satisfied by these conserved charges. \\

\paragraph{Algebra}
In this section, we compute the algebra of the conserved charges \eqref{charges} together with the condition \eqref{restconscharges}. 
As in \cite{Compere:2007in}, we write an element of the algebra as
\begin{equation}\label{key}
a=(\xi,\lambda,\Lambda)
\end{equation}
and we use the following bracket
\begin{equation}\label{key}
[a,a']_G\equiv ([\xi,\xi'],\mathsterling_\xi \lambda'-\mathsterling_{\xi'}\lambda ,\mathsterling_\xi \Lambda'-\mathsterling_{\xi'}\Lambda )\,.
\end{equation}
We define 
$
\ell_n=(\xi_n,0,0)
$ where $\xi_n= - i r n \text{e}^{in\phi}\p_r+ \text{e}^{in\phi}\p_\phi $ and by abuse of notation, we denote $p_0=(\p_u,0,0)$, $q_0=(0,\lambda,0)$ and $r_0=(0,0,\Lambda_\phi \extd \phi)$. 
The $\ell_n$ satisfy a Witt algebra while the three other symmetries commute with everything. 

Now, we consider the algebra of the charges associated with parameters $a$. It takes the form 
\begin{equation}\label{bracket}
\delta_a H_{a'}:= \{H_{a'}, H_a \} =  H_{[a,a']_G} +  K(a,a')\,
\end{equation}
where the first equality is the usual definition the Poisson bracket of charges\footnote{Note that this definition needs to be modified in the presence of non-integrable charges \cite{Barnich:2011mi}, but this is not the case here.}.

First, we compute $\delta_{\ell_m } H_{\ell_n}$, we get 
\begin{equation}\label{key0}
H_{ [\ell_n,\ell_m]_G }+ \int_0^{2\pi} \extd\phi\, \p_\phi \Big[ -\frac{1}{16\pi} u_{01} \text{e}^{i(n+m)\phi+4f_{00}  } \Big] 
\end{equation}
so the Witt algebra does not pick up a central extension, the extra term being a boundary term.

The expression $\delta_{p_0} H_{q_0}$ gives a zero contribution recalling that the zero mode of $\mathcal A$ is independant of $u$, 
\begin{equation}\label{key3}
\delta_{p_0} H_{q_0} = \frac{k_g \, X\, \lambda}{32\pi L^2} \p_u \int_0^{2\pi} \extd \phi  \mathcal A =0\,.
\end{equation}
Similarly $\delta_{a} H_{p_0}$ gives 
\begin{equation}\label{key4}
\delta_{p_0} H_{p_0}  = \frac{L}{16\pi}\,\p_u \int_0^{2\pi} \extd \phi  \, V_{00}=0 \,. 
\end{equation}
Finally,  $\delta_{a } H_{r_0}$ gives 0. 

The asymptotic symmetry algebra thus consists in a centerless Virasoro algebra supplemented by three exact charges $u(1)$ charges.  

\section{Solutions in phase space}

In the previous section, we have established a phase space including the black string solution. An interesting question is whether there exists other saddle points in the phase space. We address this question after reviewing other known classical solutions belonging to the phase space.


\subsection{Horne--Horowitz black string}

An interesting solution included in the boundary conditions is the Horne--Horowitz black string, corresponding to solution (\ref{BS4met}) with a vanishing Maxwell field. In our conventions\footnote{We relegate the details of the transposition of the original expression of \cite{Horne:1991gn} to Bondi gauge in App. \ref{appHH}.}, the Horne--Horowitz black string is 
\begin{align}
& \text{d}s_{\text{H}}^2=4\left (-r^2+r\left(M_\text{H}+\frac{Q_{\text{H}}^2}{M_{\text{H}}}\right) \right )\text{d}u^2-2 r\,\text{d}u\,\text{d}r+4Q_{\text{H}}\,r\,\text{d}u\,\text{d}\phi+r^2\,\text{d}\phi^2 \\
& \Phi = -\frac12\log \left(\frac rL\right) \\
&\left(H_{ur\phi}\right)_{{\text{H}}}=\frac{ L^2 \, Q_{\text{H}}}{ r^2}\,,
\end{align}	
and corresponds to the point in the phase space with 
\begin{equation}\label{key}
f_{00}=0\,, \quad u_{01}=2Q_{\text{H}}\,, \quad  V_{00}=\frac4L\left( M_{\text{H}}+\frac{Q_{\text{H}}^2}{M_{\text{H}}} \right) \,, \quad \Omega= \frac{2Q_{\text{H}}}{L^2}
\end{equation}
with the other functions and subleading terms set to zero. It naturally fits the general black-string solution \eqref{BS4} with 
\begin{equation}\label{key}
\alpha=0\Leftrightarrow r_-=\frac{\zeta^2\omega^2}{r_+}, \quad r_+=M_{\text{H}}, \quad \zeta=1\,, \quad \omega=Q_{\text{H}}.
\end{equation}

\subsection{Two-dimensional black hole} \label{sec:two-dimensionalbh}

Switching off the Kalb--Ramond field in the Horne--Horowitz black-string solution, \emph{\emph{i.e.}} setting $Q_{\text{H}}=\omega=0$, leads to a metric plus dilaton background 
\begin{equation}
\label{two-dimensionalBH3-n}
\text{d}s^2=  -2r \text{d}u \text{d}r-4r  (r-M_{\text{H}})  \text{d}u^2+r^2 \text{d}\phi^2,
\end{equation}
and 
\begin{equation}
\label{matterBS-n}
 \Phi = -\frac12\log \left(\frac rL\right).
\end{equation}
The metric background in \eqref{two-dimensionalBH3-n} together with the dilaton field \eqref{matterBS-n} is a remarkable solution, consisting of a decoupled free (not necessarily compact) direction $\phi$ together with the two-dimensional Lorentzian black hole \cite{Witten:1991yr}.

\subsection{Time-dependent solutions}


Returning to the equations of motion, it is possible to find more general explicit solutions to the equations of motion satisfying our boundary conditions \cite{Spindel:2018cgm}.
Indeed, the $rr$-component of Einstein's equation and the $u$-component of Maxwell's equation, read as follows:
\begin{align}\label{Err44}
&\frac1r \p_r \beta = \frac{k_g}{4L^2}\text{e}^{-4f_{00}} (\p_rA_\phi)^2+\frac1{r^2}(1-2 r \p_r f)^2 \\ \nonumber
&-r^3 \p_r^2 A_u+r^2 \p_r A_u(-3+r\p_r\beta4r\p_r f) + \p_r\p_\phi A_\phi \text{e}^{\beta}L\, r+ r^3 \p_r^2 A_\phi\,  U 
\\& 
\qquad \qquad +\p_rA_\phi (-\text{e}^{\beta+4F} \Omega\,L^3-4r\text{e}^{\beta} L\p_r f+r^3 \p_r U +r^2U(-3+r\p_r\beta+4r\p_rf))=0 \label{EOMA45}\,. 
\end{align}

Switching off $A_\phi$ turns equation \eqref{EOMA45} into a relation between $A_u$ and $\beta$. In addition $\beta$ is then given only in terms of the function $f$.

Furthermore, we restrict $f$ to be a function $f_{00}(\phi)$, motivated by the fact that only this component of $f$ appears in the charges. The solutions are
\begin{equation}
\beta=\log\left(\frac rL\right) +\beta_{00}(u,\phi) \,,\quad A_u=\alpha_{00}(u,\phi)+\alpha_{01}(u,\phi)\frac1r \,.
\end{equation}
Then, we successively solve $E_{r\phi},E_{ur},E_{\phi\phi}=0$ and ${\cal J}^r=0$, and we choose the integration constant such that the solution is compatible with the  phase space \eqref{BC}.
We get
\begin{align}\nonumber
 \beta=&\log\left(\frac rL\right) -4f_{00}(\phi)\,, \quad A_u=\frac{\alpha_{01}(u,\phi)}r\,, \\ \nonumber
 U=& 2\text{e}^{-4f_{00}(\phi)} \p_\phi f_{00}(\phi) +\frac{\Omega\,L^2}r\,,\\ \label{extraexactsol}
 V=& -\frac4L(1+ \text{e}^{-4f_{00}(\phi)} (\p_\phi f_{00} (\phi))^2)r+V_{00}(u,\phi)-(\frac{k_g}{4L^3 }\alpha_{01}(u,\phi)^2+\text{e}^{f_{00}(\phi)}\Omega^2L^3)\frac1{r}\,,
\end{align}
and also that $E_{u\phi}=0$ and the dilaton equation of motion are satisfied. 

The last unsolved equations are $E_{uu}=0$ and ${\cal J}^r=0$, being PDEs for $V_{00}$ and $\alpha_{01}$ respectively.
The latter is 
\begin{equation}\label{key}
\p_\phi^2 \alpha_{01} + \text{e}^{4f_{00}} \p_u \alpha_{01} -6\p_\phi f_{00}\p_\phi \alpha_{01}+(8(\p_\phi f_{00}) ^2-2\p_\phi^2 f_{00}) \alpha_{01}=0\,.
\end{equation}
We observe that for $f_{00}(\phi)$ being a constant, conveniently chosen to be $1/2 \log(\zeta)$, the equation becomes a heat equation whose solution is 
\begin{equation}\label{key}
\alpha_{01}=\sum \alpha_n \exp \left(\frac{n^2 u}{\zeta ^2}+i n \phi \right)\,.
\end{equation}
Also, $E_{uu}$ turns out to be a heat equation with solution 
\begin{equation}\label{key}
V_{00}=\sum V_n \exp \left(\frac{n^2 u}{\zeta ^2}+i n \phi \right)\,.
\end{equation}
The corresponding time-dependent solution now reads as \cite{Spindel:2018cgm}
\begin{subequations}\label{BSnew}
\begin{align} \nonumber
\extd s^2&=\left( -\frac{4}{\zeta ^2} r^2 + \frac{V_n \text{e}^{\frac{n^2 u}{\zeta ^2}+i n \phi }L}{\zeta ^2} r -\frac{k_g \alpha_n^2 \text{e}^{2 n \left(\frac{n u}{\zeta ^2}+i \phi \right)}}{4 \zeta ^2L^2}+2L^4\Omega^2\right)\extd u^2 -2\frac{r}{\zeta ^2}  \extd u\extd r  \\ 
& \hspace{1cm} +2 r\,\Omega \,L^2\extd u\extd \phi +r^2\extd\phi^2 \\ 
 A&=\frac{\alpha_n \text{e}^{\frac{n^2 u}{\zeta ^2}+i n \phi }}{r} \extd u \\ 
 	\Phi & =-\frac12\log \left( \frac rL\right) +\frac12 \log\zeta\,.
\end{align}
\end{subequations}
A general solution consists in a superpositions of the above modes, while the black string corresponds to $\alpha_n,V_n=0$ for $n\neq 0$. This family of solutions deserves further study; this will be addressed elsewhere.


\section{CFT Interpretation}\label{SectCFT}

As already mentioned, the charged black-string solution, Eqs. \eqref{BS4}, was reached as a double marginal deformation of the $SL(2,\mathbb{R})$ WZW model at level $k$ with $\delta c=\nicefrac{12\alpha'}{L^2}=\nicefrac{6}{k-2}$. Alternatively, in the coordinates at hand, it appears naturally as a double marginal deformation of the two-dimensional black hole (Sec. \ref{sec:two-dimensionalbh})
driven by parafermion bilinears.  
Although this discussion falls outside of our main goal in the present work, it is worth making these statements more precise, as this will help giving another perspective to the new families of solutions
\eqref{BSnew}.

\subsection{The two-dimensional black hole as a seed for deformations}

Our starting point is the general black-string solution \eqref{BS4}, described in terms of fundamental parameters $\omega$, $r_+$ and $r_-$. Consider a special locus in the parameter space, corresponding to 
\begin{equation}
\label{two-dimensionalBH3con}
r_-=\frac{\zeta^2\omega^2}{r_+}\quad \text{and}\quad \omega=0. 
\end{equation}
This configuration has neither gauge field nor Kalb--Ramond. It has dilaton (Eq. \eqref{matterBS}) and metric (as in \eqref{two-dimensionalBH3-n} with general $\zeta$)
\begin{equation}
\label{two-dimensionalBH3}
\text{d}s^2_0=  -\frac{2r}{\zeta^2} \text{d}u \text{d}r-\frac{4r}{\zeta^2}  (r-r_+)  \text{d}u^2+r^2 \text{d}\phi^2.
\end{equation}
It is useful at this stage to move to the string frame (hatted fields) defined by the following rescaling:
\begin{equation}
\label{strein}
\hat{g}_{\mu\nu}= \text{e}^{4\Phi}g_{\mu\nu}.
\end{equation}
The metric \eqref{two-dimensionalBH3} for the string background subject to  \eqref{two-dimensionalBH3con} reads now (we use \eqref{matterBS} in \eqref{strein}):
\begin{equation}
\label{two-dimensionalBH3str}
\text{d}\hat{s}^2_0= -2 L^2\left(\frac{\text{d}u \text{d}r}{r}+2\frac{r-r_+}{r}\text{d}u^2\right)+\zeta^2 L^2 \text{d}\phi^2.
\end{equation}

The metric background in \eqref{two-dimensionalBH3str} together with the dilaton field \eqref{matterBS} is a remarkable string solution, consisting of a decoupled free (not necessarily compact) boson $\phi$ together with the two-dimensional Lorentzian black hole obtained as an $\frac{SL(2, \mathbb{R})}{\mathbb{R}}$ gauged WZW model \cite{Witten:1991yr}. More precisely, Eqs. \eqref{matterBS} and \eqref{two-dimensionalBH3str}, where $L^2$ is traded with\footnote{It is customary to choose algebrized units, where $\alpha'=\nicefrac{1}{2}$.} $2k\alpha'$, provides the leading order in large $k$ of the exact conformal sigma model  $\frac{SL(2, \mathbb{R})_k}{\mathbb{R}} \times  \mathbb{R}_\phi$. The corresponding exact background fields can be found in a resummed form
for all $\nicefrac{1}{k}$ orders in Refs. \cite{BS3, Bars:1992sr}.\footnote{Notice that the free boson part $\phi$ is exact \emph{per se}.} Observe for further use that the background metric of the charged black string \eqref{BS4met} can be recast in the string frame using explicitly the gauge field $A$:
\begin{eqnarray}
\label{BSmetprimestr}
\text{d}\hat{s}^2&=&\text{d}\hat{s}^2_0+ \frac{4L^2}{r} \left(r_- \text{d}u +\omega \zeta^2 \text{d}\phi \right) \text{d}u
-\frac{k_g}{4}A^2\nonumber \\
&=&\text{d}\hat{s}^2_0+ \frac{4\omega \zeta^2 L^2}{r} \left(\frac{\omega}{r_+} \text{d}u + \text{d}\phi \right) \text{d}u+
\frac{k_g \alpha^2}{rr_+}\text{d}u^2
-\frac{k_g}{4}A^2
\label{BSmetprimestr}
\end{eqnarray}
with $A$ and $\alpha$ given in \eqref{BS4A}.

Notice finally that under the assumption \eqref{two-dimensionalBH3con} and using \eqref{dudvpm}
it is possible to trade $r$ for $v$, while keeping $\phi$, which is identical to $\psi$:
\begin{equation}
\label{uvrmzer}
v-u=\frac{1}{2}\log \frac{r-r_+}{r_+}.
\end{equation}
The two-dimensional black-hole plus free-boson metric \eqref{two-dimensionalBH3str} is then recast as 
\begin{equation}
\label{two-dimensionalBH3struv}
\text{d}\hat{s}^2_0=2k\alpha'\left( -4 \frac{\text{d}u \text{d}v}{1+\text{e}^{2(u-v)}}+\zeta^2 \text{d}\phi^2\right).
\end{equation}

The question we would like to discuss is how the charged-black-string solution is connected to the two-dimensional black-hole background in terms of exact marginal deformations in the space of conformal field theories. As opposed to WZW models, gauged WZW do not possess left and right weight-one currents that enable to build dimension-two exact marginal operators. Nevertheless, other remarkable chiral operators do exist in these coset conformal field theories, known as \emph{parafermions} \cite{BCR}. For the $\frac{SL(2, \mathbb{R})}{\mathbb{R}}$ coset, these are the Abelian non-compact parafermions, obeying 
\begin{equation}
\label{parafcons}
\partial\bar{\Psi}_\pm=0 \quad \text{and} \quad
\bar{\partial}{\Psi}_\pm=0, 
\end{equation}
and generating infinite-dimensional chiral algebras ($\partial$ and $\bar{\partial}$ refer to the world-sheet holomorphic and antiholomorphic coordinates $z$ and $\bar z$). Their semi-classical expressions in terms of the sigma-model fields $v(z,\bar z)$ and $u(z,\bar z)$ read: 
\begin{eqnarray}
\Psi_+=2\sqrt{\frac{kr_+}{r}} \partial v \, \text{e}^{2(v-i\chi)}
,&&\Psi_-=-2\sqrt{\frac{kr_+}{r}} \partial u\,  \text{e}^{-2(u-i\chi)},
\label{paraf}
\\
\bar\Psi_+=2\sqrt{\frac{kr_+}{r}} \bar\partial v\,  \text{e}^{2(v+i\chi)}
,&&
\bar\Psi_-=-2\sqrt{\frac{kr_+}{r}} \bar\partial u\,  \text{e}^{-2(u+i\chi)}
.
\label{barparaf}
\end{eqnarray}
In these expressions, $\chi$ 
%
%
is a non-local phase. Parafermions are thus non-local objects, which have non-trivial braiding properties. We will ignore this phase in our discussion.

Parafermions appear in the  expression of the energy--momentum tensor. The holomorphic component of the latter, for example, reads  (using \eqref{two-dimensionalBH3str}, \eqref{uvrmzer} and \eqref{paraf}):
\begin{equation}
\label{T}
T=\hat g_{\mu\nu}\partial x^\mu\partial x^\nu =2\alpha' \left(\Psi_+\Psi_-+k (\partial \phi)^2\right), 
\end{equation}
and similarly for the antiholomorphic one. This  expression receives quantum corrections because
parafermions have anomalous dimensions: their conformal weights are, at leading (semi-classical) order in $\nicefrac{1}{k}$, $h=1+\nicefrac{1}{2k}$ for the holomorphic ones, and $\bar h=1+\nicefrac{1}{2k}$ for the antiholomorphic ones \cite{parafZF,lykken}. 

Due to the parafermion  anomalous dimensions, left-right bilinears such as $\Psi_+\bar\Psi_+$ or $\Psi_-\bar\Psi_-$ \emph{are not }marginal $(h,\bar h)=(1,1)$ operators. However, as it was observed in \cite{PS05,PS06}, conformal composite operators based on various elementary fields make it possible to promote the parafermion bilinears onto marginal operators, by adjusting their conformal weights. 

To that end we should recall that $\frac{SL(2, \mathbb{R})}{\mathbb{R}}$ operators originate from the $SL(2, \mathbb{R})_k$-WZW affine primaries and their descendants. These can be constructed as composite operators of group elements $g(z,\bar z)\in SL(2, \mathbb{R})$. Following the gauging procedure and performing the appropriate gauge fixing, one reaches the  $\frac{SL(2, \mathbb{R})}{\mathbb{R}}$ fields. As an example, we quote several such composite operators, corresponding to lowest or highest-level $SL(2, \mathbb{R})$ representations of lowest spin (the indices refer to the left and right $SL(2, \mathbb{R})$ spin-$\nicefrac{1}{2}$ projections  -- the interested reader can find details for this construction \emph{e.g.} in \cite{PS05}):
\begin{equation}
\label{comp}
g_{++}=\text{e}^{-2u},\quad g_{--}=\text{e}^{2v},\quad g_{+-}=g_{-+}=\sqrt{\frac{ r }{r_+}}.
\end{equation}
This set  has semiclassical conformal weight $\nicefrac{1}{4k}$. Products of these operators provide further composite fields. At the semiclassical level their weights are additive, but higher-order $\nicefrac{1}{k}$ corrections usually appear.   
%

Besides the  $\frac{SL(2, \mathbb{R})}{\mathbb{R}}$ conformal operators, the free boson $\phi(z,\bar z)$ brings its own tower of conformal states: the left and right currents, $\partial \phi$ and $\bar\partial \phi$, as well as the vertex operators
\begin{equation}
\label{vertop}
V_\gamma=\text{e}^{2\gamma \zeta \phi},
\end{equation}
of conformal weight $\nicefrac{-\gamma^2}{2k}$. Notice that $\gamma$ is either real or imaginary.
 
 Let us now turn to the marginal deformations of the $\frac{SL(2, \mathbb{R})_k}{\mathbb{R}} \times  \mathbb{R}_\phi$ sigma model described in terms of background metric \eqref{two-dimensionalBH3str} and dilaton \eqref{matterBS}.  Remember that the general sigma-model action reads:
\begin{equation}\label{eq:sigma}
  S[x]=\frac{1}{4\pi\alpha'}\int \mathrm{d}z^2\,
  \big( \hat g_{\mu\nu}(x) + B_{\mu\nu}(x) \big)\partial x^\mu\, \bar{\partial} x^\nu  -
  \frac{1}{8\pi}\int \mathrm{d}z^2\,  \Phi(x) R_{(2)}.
\end{equation}
Any dimension-two operator $\mathscr{O}(z,\bar z)$, added as $\frac{1}{4\pi\alpha'}\int \mathrm{d}z^2\,\mathscr{O}(z,\bar z)$, produces a deformation $ \delta g_{\mu\nu}(x)$ and $\delta B_{\mu\nu}(x) $. Often  $\mathscr{O}(z,\bar z)$ is factorized in holomorphic/antiholomorphic pieces, but this needs not be the case, as we will see here, along the lines of  \cite{PS05,PS06}. It is appropriate to stress here that gauge fields are introduced as marginal deformations induced by Kaluza--Klein reductions.\footnote{This important issue was originally discussed in \cite{Kiritsis:1995iu} and further adapted to this general context in \cite{Israel:2004vv, Petropoulos:2005, Israel:2004cd}.}
Such a reduction brings an extra term to the world-sheet action $S[x]$:
\begin{equation}\label{eq:sigma-A}
  \frac{1}{4\pi\alpha'}\int \mathrm{d}z^2\,
  \left(A \bar J_g +  \bar A J_g +\frac{k_g}{4} J_g \bar J_g \right),
  \end{equation}
where $A=A_\mu \partial x^\mu$ is the gauge-field operator and $J_g =\partial y$ the  gauge current realized at level $k_g$ in some internal algebra after the reduction of the fourth dimension along the coordinate $y$. The extra term \eqref{eq:sigma-A} creates a gauge field $A=A_\mu \text{d}x^\mu$ and deforms the metric as 
\begin{equation}\label{eq:sigma-A-metdef}
  \hat g_{\mu\nu}(x) \to  \hat g_{\mu\nu}(x) -\frac{k_g}{4}A_\mu A_\nu. 
   \end{equation}

Deformations may or may not be integrable. In the former case, the operator survives its own perturbation and a continuous line of conformal sigma models is produced. The investigation of this property lies beyond our scope, and we will limit our presentation to exhibiting the operators which generate the black string and the charged black string starting from the free boson plus two-dimensional black-hole background.

The deformed backgrounds have vanishing beta functions, even at lowest order in $\alpha'$. This is an argument -- although not a proof -- in favour of  the corresponding marginal operators being exact (integrable). Since the black string and the charged black string are exact conformal backgrounds (by construction \cite{Detournay:2005fz, Horne:1991gn}), this argument is very strong. In the following, we will build a new family of backgrounds by exhibiting a marginal operator generating a deformation of the free boson plus two-dimensional black-hole. 
We will not prove the conformal exactness of the operator, but the finite deformation reached in this way turns out to be an extremum of \eqref{lagrangian}.

Two operators play a role in the investigation of the black string, both marginal in the semi-classical approximation:
\begin{eqnarray}
\mathscr{O}_{\Psi\Psi}&=& \frac{\Psi_-\bar  \Psi_-}{g_{++}^2},
\label{eq:O1}
\\
\label{eq:O2}
\mathscr{O}_{\Psi\phi}&=& - \frac{\Psi_-}{g_{++}g_{+-}}\sqrt{k}\bar  \partial \phi.
   \end{eqnarray}
With these operators, we can generate three marginal deformations. The starting point is the two-dimensional black hole with metric and dilaton background \eqref{matterBS} and \eqref{two-dimensionalBH3str} (in the string frame). 
\begin{enumerate}
\item Adding 
\begin{equation}\label{eq:def1}
2\alpha' \frac{\delta r_+}{r_+} \mathscr{O}_{\Psi\Psi}
   \end{equation}
to the world-sheet Lagrangian\footnote{Remember that $L^2\approx 2 \alpha' k$.}
deforms only the metric, by shifting the horizon $r_+$ to $r_+ + \delta r_+$.
\item Adding instead 
\begin{equation}\label{eq:def2}
2\alpha' \zeta^2\frac{\omega}{r_+} \left(\frac{\omega}{r_+} \mathscr{O}_{\Psi\Psi}+2 \mathscr{O}_{\Psi\phi}\right)
   \end{equation}
deforms both the metric and the Kalb--Ramond field $B$, and allows to recover the neutral black string  (in string frame) \eqref{BSmetprimestr} with dilaton \eqref{matterBS}  and Kalb--Ramond field \eqref{BS4B}. The gauge field  \eqref{BS4A}  vanishes so that $r_-$ takes the value displayed in \eqref{two-dimensionalBH3con} -- with non-zero $\omega$ though. 
\item Finally the deformation\footnote{The first term is unnecessary for the purpose of switching on the electric field. It is meant to keep $r_+$ unaltered, which would have been affected otherwise.} 
\begin{equation}\label{eq:def3}
\frac{ k_g \alpha^2}{4kr_+^2}\mathscr{O}_{\Psi\Psi} +A_{\Psi}\bar \partial y +\bar A_{\Psi} \partial y +\frac{k_g}{4} \partial y  \bar \partial y 
   \end{equation}
with
\begin{equation}\label{eq:def3A}
A_{\Psi}=- \frac{\alpha}{r_+\sqrt{k}}
 \frac{\Psi_-}{g_{++}g_{+-}},
   \end{equation}
applied to the neutral black string, switches on the gauge field 
\begin{equation}\label{eq:def3Ag}
A= \frac{2\alpha}{r}\text{d}u,
   \end{equation}
and the string background is given in  \eqref{BSmetprimestr} (in string frame) with \eqref{matterBS}, \eqref{BS4A} and \eqref{BS4B}. Hence
the parameters $\alpha$ and $r_-$ are related through the identification of 
\begin{equation}
r_- = \frac{\alpha^2k_g}{4L^2r_+}
+\frac{\zeta^2 \omega^2}{r_+}.
\end{equation}

\end{enumerate}

\subsection{New deformations beyond the charged black string}

\paragraph{Generic case}

We will now propose new deformations of the charged black string. These correspond to the patterns 1. and 3. met above, with the marginal operator $\mathscr{O}_{\Psi\Psi}$ in \eqref{eq:def1} traded for 
\begin{equation}
\mathscr{O}_{\gamma}=V_\gamma\,  g_{++}^{2\gamma^2} \, \frac{\Psi_-\bar  \Psi_-}{g_{++}^2},
\label{eq:O1new}
   \end{equation}
and the operator $A_\Psi$  in \eqref{eq:def3} replaced with 
\begin{equation}\label{eq:def3Anew}
A_{\gamma}=- \frac{\beta }{r_+\sqrt{k}}V_\gamma \, g_{++}^{2\gamma^2} \, 
 \frac{\Psi_-}{g_{++}g_{+-}},
   \end{equation}
   where $\beta$ is an arbitrary constant.
The composite operator $V_\gamma \, g_{++}^{2\gamma^2} $ has indeed dimension zero for any real or imaginary $\gamma$ (see the weights in \eqref{comp} and \eqref{vertop}). We have no proof that the $\gamma$-deformations are exactly marginal for $\gamma\neq 0$. It turns out that the corresponding deformed configurations solve the low-energy string equations of motion (as we will see soon) and this suggests that the operators at hand might be exactly marginal, possibly after correcting them with higher order $\nicefrac{1}{k}$ corrections. 

When acting on the charged black string, the operator \eqref{eq:O1new} induces a metric deformation only, given in the string frame by 
\begin{equation}
\delta\text{d}\hat s^2\propto \frac{1}{r}\text{e}^{2\gamma (\zeta\phi-2\gamma u)}\text{d}u^2.
\label{eq:O1newmetdef}
   \end{equation}
The operator \eqref{eq:def3Anew}, within the combination \eqref{eq:sigma-A} acting on the neutral black string deforms both the metric and the gauge field as follows:
\begin{eqnarray}
A&=& \frac{2\beta }{r}\text{e}^{2\gamma (\zeta\phi-2\gamma u)} \text{d}u,
\label{eq:AnewAdef}
\\
\delta\text{d}\hat s^2&=&-\frac{ k_g \beta^2}{r^2}\text{e}^{4\gamma (\zeta\phi-2\gamma u)}\text{d}u^2.
\label{eq:Anewmetdef}
   \end{eqnarray}
where $\beta$ is an arbitrary constant deformation parameter.

Putting the above transformations together, we finally find the following exact extremum of \eqref{lagrangian}, expressed in the Einstein frame:
\begin{equation}
\label{BSmetprimenew}
\text{d}s^2= -\frac{2r}{\zeta^2} \text{d}u \text{d}r-\frac{4r}{\zeta^2}  \left(r-r_+-\frac{\zeta^2\omega^2}{r_+}
\right)  \text{d}u^2-\frac{ k_g\beta^2 }{\zeta^2L^2}\text{e}^{4\gamma (\zeta\phi-2\gamma u)}\text{d}u^2+4 \omega r \text{d}\phi\text{d}u+r^2 \text{d}\phi^2
\end{equation}
with dilaton, Kalb--Ramond field and gauge field given in \eqref{matterBS},  \eqref{BS4A} and \eqref{eq:AnewAdef}. The scalar curvature of the background at hand reads:
\begin{equation}
\label{BScurvprimenew}
R=-
\frac{2\zeta ^2}{r^2 } \left(4 +\frac{2 r_+}{r}+\frac{2\zeta ^2 \omega ^2}{r r_+} -\frac{5 \zeta ^2 \omega ^2}{r^2}- \frac{k_g \beta ^2}{r^2 L^2} \text{e}^{4 \gamma  (\zeta  \phi -2 \gamma  u)}\right). 
\end{equation}

As a closing remark, we would like to stress that the success in generating new families of exact lowest-order string solutions suggests that the parafermionic operators introduced so far are indeed exact, up to higher-order $\nicefrac{1}{k}$ corrections. 
We should also notice that all of our deformations can be repeated by trading $u$ for $-v$. This will provide a kind of mirror set of solutions.  
Combining deformations is not allowed, unless the corresponding operators commute. This is usually not the case and checking it would require to determine operator product expansions, which is far beyond our motivations here.

\paragraph{Periodic $\pmb{\phi}$}

This is the framework discussed all over the present work, and for that we need $\gamma = \text{i}\tilde \gamma $. The deformed fields read now:
\begin{eqnarray}
A&=& \frac{2\beta }{r}\text{e}^{4\tilde\gamma^2 u} \cos (2 \tilde \gamma \zeta \phi) \text{d}u,
\label{eq:AnewAdef-im}
\\
\text{d}s^2&=& -\frac{2r}{\zeta^2} \text{d}u \text{d}r-\frac{4r}{\zeta^2}  \left(r-r_+-\frac{\zeta^2\omega^2}{r_+}
\right)  \text{d}u^2
\nonumber
\\
&&-\frac{ k_g\beta^2 }{\zeta^2L^2} \text{e}^{8\tilde\gamma^2 u} \cos^2 (2 \tilde \gamma \zeta \phi) 
\text{d}u^2+4 \omega r \text{d}\phi\text{d}u+r^2 \text{d}\phi^2,
\label{eq:Anewmetdef-im}
   \end{eqnarray}
together with  \eqref{matterBS} and \eqref{BS4A}. 
The scalar curvature reads:
\begin{equation}
\label{BScurvprimenew}
R=-
\frac{2\zeta ^2}{r^2 } \left(4 +\frac{2 r_+}{r}+\frac{2\zeta ^2 \omega ^2}{r r_+} -\frac{5 \zeta ^2 \omega ^2}{r^2}- \frac{k_g \beta ^2}{r^2 L^2}  \text{e}^{8\tilde\gamma^2 u} \cos^2 (2 \tilde \gamma \zeta \phi) \right). 
\end{equation}
Solution \eqref{eq:Anewmetdef-im} is precisely the saddle point \eqref{BSnew} found in our general phase-space analysis, with the identification $2 \tilde \gamma \zeta=n$, and after trading the imaginary exponentials for trigonometric functions.

\section{Thermodynamics}
In this section, we focus on the thermodynamic properties of the black string solution \eqref{BS4}. 
This solution has two horizons, at $r_\pm$, with Bekenstein--Hawking entropies
\begin{equation}\label{S+}
	S_\pm=\frac\pi2 \,r_\pm
\end{equation}
and four exact symmetries, $\p_u,\p_\phi,\lambda$ and $\extd \phi$, with four corresponding conserved charges. 

First of all, we derive geometrically the thermodynamic potentials.  
The angular velocity and Hawking temperature at the outer horizon can be easily determined:
\begin{equation}
	\Omega_+ =\frac{2\omega} {L\,r_+}\,,\qquad T_+ =\frac{(r_+-r_-)}{L\,\pi\, r_+} \, . 
\end{equation}   
The electric potential is defined by \cite{Gao:2001ut}
\begin{equation}
	\Phi_A^+=\left. (\chi \cdot A)\right|_{r_+}=\frac{ 4\sqrt{ r_- r_+ -\zeta^2 \omega^2 }}{\sqrt{k_g} r_+},
\end{equation}
where $\chi$ is the generator of the horizon, namely $\nicefrac{1}{L} \p_u-\Omega_+\p_\phi$. For the $B$-field, the potential is given by \cite{Copsey:2005se} (the general expression for a p-form can be found in \cite{Compere:2007vx})
\begin{equation}
	\Phi_B^+=\frac {2 \omega\,L^2\,\zeta^2}{r_+}\,.
\end{equation}

The exact conserved gravitational charges (mass and angular momentum) associated to the black string solution can readily be obtained from the general expression \eqref{intdk} as
\begin{equation}\label{key}
	\delta M =\delta H_{1/L\p_u}=\frac{(\delta r_+ +\delta r_-)}{2 L}\, ,\qquad -\delta J=\delta H_{\p_\phi}=-\frac{\zeta}{4}(\zeta\,\delta \omega +2\delta \zeta\,\omega)\,,
\end{equation}
where the variation of a given solution is taken with respect to its four parameters $(r_+,r_-,\omega,\zeta)$.
The finite expressions for the charges are thus given by
\begin{equation}\label{key1}
	M= \frac{(r_+ +r_-)}{2L}\, ,\qquad 
	J =\frac{\omega\,\zeta^2}{4}\,.
\end{equation}
The variation of the electric charge (\emph{i.e.} the charge associated to the gauge parameter $\lambda=1$) is given by
\begin{equation}\label{key2}
	\delta Q_A:=\delta H_{\lambda=1}=\frac{\sqrt{k_g}}{8 L} \frac{-2\omega \, \zeta(\omega\,\delta \zeta+\zeta\,\delta\omega)+ (r_+ \delta r_-+r_-\delta r_+) }{\sqrt{- \omega^2 \,\zeta^2+ r_- r_+ }}
\end{equation}
and can easily be integrated to get the electric charge
\begin{equation}\label{key2}
	Q_A=\frac{\sqrt{k_g}}{ 4L} \sqrt{- \omega^2 \,\zeta^2+ r_- r_+}\,.
\end{equation}
Similarly, the variation of the $B$-field charge is defined as
\begin{equation}\label{key3}
	\delta Q_B:=\delta H_{\Lambda=1/Ld\phi}=\frac{\delta\omega}{4L^3}\,,
\end{equation}
yielding
\begin{equation}\label{key}
	Q_B=\frac{\omega}{4L^3}\,. 
\end{equation}

The first law of thermodynamics is a direct consequence of the vanishing divergence of $k_{(\chi,0,0)}(\Psi,\delta \Psi)$ \cite{Wald:1993nt, Iyer:1994ys, Compere:2007vx, Compere:2006my}, implying in particular that
\begin{equation}\label{intdk2}
	\int_{S_\infty} k_{(\chi,0,0)}(\Psi,\delta \Psi)= \int_{\Sigma_+} k_{(\chi,0,0)}(\Psi,\delta \Psi),
\end{equation}
where $S_\infty$ is the circle at infinity, $\Sigma_+$ the outer horizon and $\chi$ its generator.
The l.h.s. accounts for the terms
\begin{equation}
	\delta M - \Omega_+ \delta J \,,
\end{equation}
while the r.h.s. decomposes into various contributions:
\begin{align}
	&\int_{\Sigma_+} k^{grav}  =  \frac{(r_+-r_-)\delta r_+}{4 r_+} = T_+ \delta S_+\,,\qquad  \int_{\Sigma_+} k^B = \frac{ \omega\delta \omega}{4 r_+} = \Phi_B^+ \delta Q_B \,,\\
	& \int_{\Sigma_+} k^A = \frac{r_+\delta r_-+r_-\delta r_+-2 \omega\delta \omega}{4 r_+} =\Phi_A^+ \delta Q_A\,,\qquad \int_{\Sigma_+}  k^\Phi = 0\,.
\end{align}
The first contribution turns into the variation of the entropy times the Hawking temperature, while the second and third ones give the $A$,$B$-potential times the $A$,$B$-charges. 
Equation \eqref{intdk2} then turns into the the first law of thermodynamics at the outer horizon,
\begin{equation}
	\delta M = T_+ \delta S_+ + \Omega_+ \delta J +\Phi_A^+ \delta H_A+\Phi_B^+ \delta H_B \,. 
\end{equation}

An interesting observation is that the Smarr formula for the black string reads as 
\begin{equation}\label{MBS}
	M=T\,S+\Omega \, J+\Phi_A\,Q_A+\Phi_B\,Q_B\,. 
\end{equation}
This leads to a vanishing Gibbs free energy, defined by Legendre transformations of the mass for all variables,
\begin{equation}\label{key}
	G(T,\Omega,\Phi_A,\Phi_B)=0\,.
\end{equation}
Thus, it enforces a relation among the four potentials. Indeed, they are related as
\begin{equation}\label{relpotential}
	\Phi_B= \frac{4L}{\Omega}\left( 1-L\, \pi  \, T -\frac{k_g \Phi_A^2}{16} \right) \,. 
\end{equation}
This is also realized in the fact that the potentials only depend on the following combinations of black string parameters $\frac{r_-}{r_+},\frac{\omega}{r_+},\zeta$. 

The property \eqref{MBS} is satisfied for systems with a homogeneous scaling in their extensive variables, which 
is usually not the case for black objects, see for example the table p5 of \cite{Dolan:2014jva}. BTZ black holes satisfy $M=\frac12 T\, S+\Omega \, J$ and therefore have a non-vanishing Gibbs free energy. More general black holes have more complicated Smarr-type relations (see for instance Sec.
IV of \cite{Caldarelli:1999xj} for the counterpart in Kerr-Newman-AdS black holes) and generically non-vanishing free energies as well. This feature clearly deserves more attention and it is left for future works. In particular, it would be interesting to confirm the vanishing of the free energy by an direct evaluation of the (appropriately regularized) on-shell action.


We also note that the variation of $M$, seen as a function of $M(S,J,Q_A,Q_B)$, can be written as
\begin{equation}\label{firstlaw}
	\delta M=\left (\frac{1}{\pi  \,L}-\frac{4\pi \,L\, \tilde{Q}}{S^2}\right ) \delta S + \frac{4\pi\,L}{S} \delta \tilde Q
\end{equation}
with the effective charge \begin{equation}\label{QT}
	\tilde Q=L\, J \,Q_B+\frac{Q_A^2}{k_g}\,.
\end{equation}

Finally, the entropy can be written in terms of the charges as
\begin{equation}\label{entropyf}
	S= \frac{\pi\,L}2 \left(M+\sqrt{M^2-16 \, \tilde Q}\right)\,.
\end{equation}
This bears some interesting similarities with the entropy of a generic Kerr black hole.
Reproducing this entropy from the symmetries suggested by the asymptotic symmetry analysis is also left as an open question. 

We close this section with a short comment on properties of the inner horizon of the black string solution. It has been observed over the years that inner horizons seemingly enjoy thermodynamic properties similar to the ones at the outer horizon \cite{Castro:2012av,Chen:2012mh,Larsen:1997ge,Detournay:2012ug}. On the one hand, the product of outer and inner horizon entropies appears to depend only on the quantized charges of the theory and is independent of the mass. 
On the other hand, inner horizons satisfy their own first law. These two observations can easily be verified to hold for the black string solution. Indeed, from (\ref{S+}), one gets
\begin{equation} \label{innerE}
	S_+ S_- = 4\pi^2\,L^2\, \tilde Q \,.
\end{equation}
Then, it can easily be verified that 
\begin{equation}
	\delta M = -T_- \delta S_- + \Omega_- \delta J +\Phi_A^- \delta Q_A+\Phi_B^- \delta Q_B \,
\end{equation} 
with
\begin{equation}
	S_-=\frac{\pi \,r_-}{2}\,,\quad \Phi_A^-=\frac{ 4\sqrt{ r_- r_+ -\zeta^2 \omega^2 }}{\sqrt{k_g} r_-}\,,\qquad \Phi_B^-=\frac {2 \omega\,L^2\,\zeta^2}{r_+}\,.
\end{equation}

Note that \eqref{firstlaw}, \eqref{entropyf} and \eqref{innerE} seem to suggest that while $Q_A$ and $Q_B$ can be varied independently, only a particular combination of them (in the form of $\tilde Q$) appears to be physical and is reflected in the thermodynamical properties of the system. 


\section{Outlook}
The original objective of our work was to determine a consistent phase space containing three-dimensional black string solutions, determine its asymptotic symmetries, and explore whether the latter could provide a preliminary explanation of their thermodynamics, in the spirit of \cite{Strominger:1997eq} for BTZ black holes.
One hope one might have had was to find a BMS$_3$ algebra, and reproduce, for a black-hole-like object, their entropy through a BMS-Cardy counting in three-dimensional-asymptotically flat spaces as has been done for cosmological spacetimes \cite{Barnich:2012xq,Bagchi:2012xr}. Our results are captured in Eqs. (\ref{BC}), \eqref{keytext} and (\ref{charges}): the asymptotic symmetry group consists in a centerless chiral Virasoro algebra, supplemented by three commuting $u(1)$ charges. We furthermore identified in our boundary conditions various solutions, both old (Horne--Horowitz black string and its generalization, two-dimensional Witten black hole) and new (time-dependent), which we interpreted as marginal deformation of the two-dimensional black-hole worldsheet theory. We finally discussed various thermodynamical properties of the general black string.

We close by listing various questions raised in the course of our analysis, and possible future directions.
\begin{itemize}
\item {\it Phase space content}  

We have not systematically explored the content of our set of boundary conditions (\ref{BC}). In particular, our black string solutions \eqref{BS4} have functionally dependent chemical potentials, see \eqref{relpotential}. Do more general black string solutions, with 4 independent potentials, exist? 

\item {\it More general boundary conditions and symmetries} 

Determining boundary conditions encompassing a given set of ``zero-mode''  solutions is a notoriously difficult task, especially when not knowing what the answer should be. Also, once a solution to the problem is found, it is not guaranteed to be unique. This is illustrated in the classic AdS$_3$ setup by the discovery, 25 years after the Brown--Henneaux boundary conditions \cite{Brown:1986nw} of a whole zoo of alternative boundary conditions (see \cite{Grumiller:2016pq} for a summary). The entropy formula (\ref{entropyf}) is reminiscent of that of a Warped CFT \cite{Hofman:2011zj,Detournay:2012pc}, appearing namely in gravity with boundary conditions with asymptotic symmetries consisting in the semi-direct product of a Virasoro and an affine $u(1)$ algebra. Do boundary conditions allowing for these symmetries exist?

\item {\it Frames} 

We have here mostly taken a relativists' point of view by working with the metric in Einstein frame (except in Sec.~\ref{SectCFT}). In most contexts, this does not make a difference because the dilaton is constant. The situation is different here. In \cite{Horne:1991gn}, the global structure of the string frame Horne--Horowitz black string was described, and shown to share similarities with the Reissner--Nordstr\o m black hole, where it was dubbed ``asymptotically flat'' . It would therefore be interesting to reproduce our analysis in the string frame.

\item {\it Near-horizon symmetries}

  The study of near-horizon symmetries as a handle to understand black-hole entropy underwent recently a renewed interest with the soft hair proposal of Hawking, Perry and Strominger \cite{Hawking:2016msc}. It would be interesting to study these for the black string solution along the lines of \cite{Donnay:2015abr, Afshar:2015wjm, Afshar:2016wfy}. Preliminary results (in string frame) have appeared for the extremal black string in \cite{Kaloper:1998vw}.

\item {\it $\alpha'$-corrections} 

The solutions presented in this work have the interesting feature of being the target space of exact string theory backgrounds. The background fields extracted from the string worldsheet action are however, generically, only valid to lowest orders in $\alpha'$ and need to be corrected. For Witten's two-dimensional black hole and the Horne--Horowitz black string, this was done in \cite{Dijkgraaf:1991ba, Bars:1992sr, Sfetsos:1992yi}. Do these corrections modify the asymptotic behaviour of the solutions and the corresponding boundary conditions?

\item {\it Asymptotic T-duality} 

Horne and Horowitz showed that the three-dimensional black string and BTZ black holes can be mapped on each other using a duality transformation \cite{Horowitz:1993jc}. As such, they might correspond to equivalent worldsheet CFTs, and a string propagating on either of the backgrounds might not be able to distinguish between them. This might sound a bit puzzling as the solutions have rather different asymptotic behaviors. How to reconcile the fact that ``equivalent''  solutions from the string theory viewpoint could possibly have rather different asymptotic symmetry groups? How do Brown--Henneaux boundary conditions map under duality?

\end{itemize}

We hope that this work has allowed to set the stage for returning to these questions in a near future.

\section*{Acknowledgements}
This work started as a collaboration with Philippe Spindel whose contributions to many aspects of it are warmly acknowledged here. His considerations on the same topic have recently appeared in \cite{Spindel:2018cgm}.
We would like to thank Alejandra Castro, Geoffrey Compere, Gaston Giribet, Daniel Grumiller, Diego Hofman, Gary Horowitz, Gim Seng Ng and Ricardo Troncoso for useful discussions. St\'ephane Detournay thanks the hospitality of the Galileo Galilei Institute for Theoretical Physics during the period 10--14 April 2017. C\'eline Zwikel and St\'ephane Detournay thank the hospitality of Ecole Polytechnique, Palaiseau, where parts of this work have been realized. Marios Petropoulos thanks the Universit\'e Libre de Bruxelles for hospitality and financial support during his numerous visits. 
St\'ephane Detournay is a Research Associate of the Fonds de la Recherche Scientifique F.R.S.-FNRS (Belgium). He is also supported  in part by the ARC grant ``Holography, Gauge Theories and Quantum Gravity Building models of quantum black holes'', by IISN -- Belgium (convention 4.4503.15) and benefited from the support of the Solvay Family. 
C\'eline Zwikel was a research fellow of ``Fonds pour la Formation à la Recherche dans l'Industrie et dans l'Agriculture"-FRIA Belgium. She thanks the Galileo Galilei Institute for Theoretical Physics (GGI) for the hospitality during the completion of this work, within the program ``New Developments in AdS3/CFT2 Holography''  and also thanks INFN as well as the ACRI (Associazione di Fondazioni e di Casse di Risparmio S.p.a.) for partial support. 
Her work is supported by the Austrian Science Fund (FWF), project P 30822. 
This work was partly funded by the ANR-16-CE31-0004 contract ``Black-dS-String'' (France).

\appendix

\boldmath
\section{The original solution in original coordinates}
\label{AppBS}
\unboldmath

The charged black string of \cite{Detournay:2005fz} is an exact background, reached by a double marginal deformation of the $SL (2, \mathbb{R})$ sigma model. Its background fields read: 
\begin{equation}
\text{d}s^2=\frac{r^2\text{d}r^2}{4\zeta^2
	L^2 \Delta(r)}-4\left(\frac{\Delta(r)}{\zeta^2}-\frac{\omega^2}{L^2}\right) \text{d}t^2+4 \frac {\omega r}{L^2} \text{d}t  \text{d}x+\frac{r^2}{L^2} \text{d}x^2,
\label{BSmetric1}
\end{equation}
and 
\begin{align}\label{matterBSapp}
& \Phi=-\frac12 \log\frac{r}{\zeta L } ,\\ 
& A= \frac{4}{r}\sqrt{\frac{r_+ r_- -\zeta^2 \omega^2}{k_g} }  \text{d}t, \label{origA}
\\
& B=\frac{2\omega \zeta^2}r \text{d}t\wedge \text{d}x , \label{origB}
\end{align}
where 
\begin{equation}
\Delta(r)=\frac{(r-r_+)(r-r_-)}{L^2}.
\end{equation}

The solution under consideration exhibits a genuine timelike singularity at $r=0$, hidden behind two horizons located at $r=r_\pm$. In the asymptotic, large-$r$ region, the metric behaves like 
\begin{equation}
\text{d}s^2\approx \frac{r^2}{L^2}\left(-\frac{4}{\zeta^2}\text{d}t^2+\text{d}x^2\right)+\frac{1}{4\zeta^2 }\text{d}r^2.\label{BSasymptmet}
\end{equation}
In order to get more insight and ensure regularity at the horizons, it is useful to move to Bondi coordinates.  We define
\begin{equation}
\frac{\text{d}t}{L}= \text{d}u +\frac {r \text{d}r}{4 L^2 \Delta(r)} \quad\mbox{and}\quad  \frac{\text{d}x}{L}=\text{d}\phi- \frac{\omega \text{d}r}{2L^2 \Delta(r)} \label{dtdupm}
\end{equation}
so that the metric \eqref{BSmetric1} becomes\footnote{This resembles the metric considered in Ref. \cite{Barnich:2013sxa}, with slightly different falloff though.}
\eqref{BS4met}, whereas the background fields \eqref{origA} and \eqref{origB} give
\eqref{BS4A} and \eqref{BS4B}. One can also trade advanced for retarded time, introducing new coordinates $v$ and $\psi$ as: 
\begin{equation}
v-u =\int\frac { r\text{d}r }{2 L^2 \Delta(r)},\quad 
\phi-\psi =\int\frac {\omega\text{d}r}{ L^2 \Delta(r)} .
\label{dudvpm}
\end{equation}
On obtains thus
\begin{equation}
\label{BSmet}
\text{d}s^2= \frac{2r}{\zeta^2}  \text{d}v \text{d}r+4\left(\omega^2
- \frac{L^2}{\zeta^2} \Delta(r) \right) \text{d}v^2+4\omega r  \text{d}v\text{d}\psi
+r^2 \text{d}\psi^2 ,
\end{equation}
whereas 
\begin{align}\label{matterBSefv}
& A= \frac {4L}{r}\sqrt{\frac{r_+ r_--\zeta^2\omega^2}{k_g} }  \text{d}v, \\
& B=\frac{2\omega\zeta^2L^2}r \text{d}v\wedge \text{d}\psi  . \label{matterBS3efv}
\end{align}

\boldmath
\section{Equations of motion in Bondi gauge}
\label{appEOM}
\unboldmath

As shown in App. A.3 of \cite{Barnich:2015jua} based on \cite{Winicour85}, the electromagnetic and gravitational Bianchi identities imply a hierarchy in the equations of motion to be solved, suggesting to first solve the four ``main'' equations $E_{rr}=0$, ${\cal J}^{u} = 0$, $E_{r\phi}=0$ and $E_{ru}=0$. Indeed, from an ansatz for $f$ and $A_\phi$, it is possible to determine $A_u,U,V$. 

We take the following ansatz for $A_\phi$ and $f$
\begin{align}
&A_\phi(u,r,\phi)= a_{00} (u,\phi)+ a_{11}  (u,\phi)\frac {\log \left( \frac rL \right) }r + a_{01} (u,\phi) \frac1r +O\left( \frac {\log^2\left( \frac rL \right)} {r^2}\right) \\
& f (u,r,\phi)= f_{00} (u,\phi)+ f_{11}  (u,\phi)\frac {\log \left( \frac rL \right) }r + f_{01} (u,\phi) \frac1r +O\left( \frac {\log^2\left( \frac rL \right)} {r^2}\right) \,,
\end{align}
and we derive the following asymptotic behaviors solving the main equation of motion
\begin{subequations}\label{FOEOMApp}
	\begin{align}
	\beta &= \log \left( \frac rL \right) + \beta_{00} + \beta_{11}\frac {\log \left( \frac rL \right) }r  + \beta_{01} \frac{1}{r} + O\left (\frac{\log^2 \left( \frac rL \right) }{r^2}\right )\\
	U &= U_{10} \log \left( \frac rL \right) + U_{00} + U_{21} \frac{\log^2 \left( \frac rL \right) }{r} + U_{11} \frac{\log \left( \frac rL \right) }{r}  + U_{01} \frac{1}{r} + O\left (\frac{\log^3 \left( \frac rL \right) }{r^2}\right )\\
	A_u &=  \alpha_{00} +  \alpha_{21} \frac{\log^2 \left( \frac rL \right) }{r} + \alpha_{11} \frac{\log \left( \frac rL \right) }{r}  + \alpha_{01} \frac{1}{r} + O\left (\frac{\log^3 \left( \frac rL \right)}{r^2}\right ) \\
\nonumber
	V & =  \bar V_{11}\, r \,\log \left( \frac rL \right)+ \bar V_{01}\, r  + V_{30} \log^3 \left( \frac rL \right)  + V_{20}\log^2 \left( \frac rL \right)+ V_{10}\log \left( \frac rL \right)\\ & \hspace{1cm}+ V_{00}   +   O\left (\frac{\log^4 \left( \frac rL \right) }{r}\right )
	\end{align}
\end{subequations}
where all symbols $\beta_{ij}$, $U_{ij}$, $\alpha_{ij}$ and $V_{ij}$ are functions of $(u,\phi)$, expressed in terms of $a_{ij}$, $f_{ij}$ and the arbitrary functions $n_{00}$ and $m_{00}$, except $\beta_{00}$, $U_{00}$, $\alpha_{00}$ and $V_{00}$ which are arbitrary so far.
For instance:
\begin{subequations}\label{F0det}
	\begin{align}
	&\beta_{01} = - 4 f_{01}, \quad \beta_{11} = - 4 f_{11}\,, \\ 
	\nonumber &\\ \nonumber
	&U_{10} = \text{e}^{\beta_{00}} \partial_\phi (4 f_{00} + \beta_{00}) \\ \nonumber
	&U_{11} =-4L \text{e}^{\beta_{00}} \left( 2f_{01} \p_\phi f_{00} - \p_\phi f_{01}-f_{11}\p_\phi(4 f_{00}+\beta_{00})\right) \\ \label{UappA}
	&U_{21}=	-2\text{e}^{\beta_{00}}\left( 2f_{11} \p_\phi f_{00} - \p_\phi f_{11}\right) 	\\
		\nonumber &\\ \nonumber
	&\alpha_{21}=\frac12 \text{e}^{\beta_{00}} \left( \p_\phi a_{11} + 2 a_{11}  \p_\phi (2f_{00}+\beta_{00})  \right) 	\\	
	&\alpha_{11}=\text{e}^{\beta_{00}} \left( \p_\phi a_{01}+a_{01} \p_\phi \beta_{00} \right) + a_{11} \left(U_{00} +\text{e}^{\beta_{00}} \p_\phi (4f_{00}+\beta_{00})  \right) \label{alphaappA}  	 
	\end{align}
	\begin{align}	\nonumber
	&\bar V_{11} = \frac2L \text{e}^{\beta_{00}} \left(\partial_\phi \beta_{00} (\partial_\phi (4 f_{00} + \beta_{00}) ) + \partial_\phi^2 (4 f_{00} + \beta_{00})) \right) \\ \nonumber
	&\bar V_{01} = \frac2L \partial_\phi U_{00} -\frac{4}{L} \text{e}^{\beta_{00}} (\text{e}^{4 f_{00}} - 3 (\partial_\phi f_{00})^2 + \p_\phi^2 f_{00} )\\ \nonumber
	& V_{30}=\frac{2}{3L} \text{e}^{\beta_{00}} \Big(-6 \p_\phi f_{00} \p_\phi f_{11}+\left(8 \p_\phi (f_{00})^2-2 \p_\phi^2 f_{00}\right)  f_{11}+\p_\phi^2 f_{11}\Big)\\ \nonumber
	& V_{20}=	\frac2L \text{e}^{\beta_{00}} \Big[ -6\p_\phi f_{00}\p_\phi f_{01}+8 \p_\phi f_{00} \p_\phi f_{11} 
	+2 \p_\phi f_{11}\p_\phi \beta_{00}+f_{01}(8(\p_\phi f_{00})^2-2\p^2_\phi  f_{00}	)  \Big] \\ \nonumber
	& V_{10}= - \frac{4}L  \text{e}^{\beta_{00}} \Big[-\left(4 \p_\phi f_{00}+\p_\phi \beta_{00}\right) \left(\p_\phi f_{01}+\p_\phi f_{11}\right) +f_{01}\left((4 \p_\phi f_{00})^2+2 \p_\phi^2 f_{00}+\p_\phi^2 \beta_{00}\right)\\ \nonumber
	& \qquad \qquad\qquad \quad - f_{11} \left(2 \left(2 \p_\phi f_{00}+\p_\phi \beta_{00}\right) \left(4 \p_\phi f_{00}+\p_\phi \beta_{00}\right)+4 \p_\phi^2 f_{00}+\p_\phi^2 \beta_{00}\right)\Big]\\\nonumber
	&  \qquad \qquad- U_{01} \left(4 \p_\phi f_{00}+\p_\phi \beta_{00}\right)+\p_\phi U_{01}  \\
	& \qquad \qquad+ f_{11}(-20 \,(\p_\phi f_{00})^2-4\p_\phi f_{00} \p_\phi \beta_{00}+2\p^2_\phi  f_{00})+\p^2_\phi f_{01}.  \label{VappA} 
	\end{align}
\end{subequations}

So far, we only took into account the main equations of motion. At this stage, we can start exploring the remaining ones. 
The equation for the dilaton at leading, $O(\log r/r^2)$, order is 
\begin{equation}\label{key}
\p_\phi T+T^2-2\, T\,\p_\phi f_{00} =0 \quad \text{ with } T:= \p_\phi(4 f_{00}+\beta_{00})\,.
\end{equation}
This equation does not have a $\phi$-periodic solution\footnote{For $T\neq0$, the solution is of the form $T= \text{e}^{2f_{00}}/(K(u)-\int^\phi \text{e}^{2f_{00}}) $. The $\phi$-periodicity requires $\int^0 \text{e}^{2f_{00}}=\int^{2\pi} \text{e}^{2f_{00}}$ which is impossible as the exponential is always positive.} expect for $T=0$. 
This leads to 
\begin{equation}
\beta_{00} = - 4 f_{00} + b_0 (u). 
\end{equation}
This in particular kills the most leading components of $V$ and $U$. 
Using the expressions for $\beta$, the next leading non-vanishing term of $E_\Phi = 0$ (order $1/r^2)$ yields
\begin{equation}
\partial_\phi U_{00} + 2 U_{00} \partial_\phi f_{00} - 2 \partial_u f_{00} + 2 \text{e}^{\beta_{00}} ( 2 (\partial_\phi f_{00})^2 - \partial_\phi^2 f_{00}) = 0\,.
\end{equation}
This is solved by
\begin{equation}\label{U00}
U_{00}=U_0(u) \text{e}^{-2 f_{00}}+2 \text{e}^{\beta_{00}}\p_\phi f_{00}+\text{e}^{-2f_{00}} \p_u F\,,\quad \text{ where } F=\int_0^\phi \text{e}^{2f_{00}(u,\theta)}d\theta \,.
\end{equation}
In order for $U_{00}$ to be periodic, the function $f_{00}$ is required to satisfy $\p_u \int^{2\pi}_0\text{e}^{2f_{00}(u,\theta)}d\theta =0$. 
Next, ${\cal J}^{\phi} = 0$ implies at leading order $O(1/r^4)$ that the leading component of the gauge field is pure gauge:
\begin{equation}\label{dadalpha}
\partial_u a_{00} = \partial_\phi \alpha_{00} \,.
\end{equation}

We have reached the expansion presented in the main part \eqref{FOEOM}. 
We didn't go further in the resolution of the equation of motions as the remaining equations involve either subleading components or non linear partial derivative equations. 


\boldmath
\section{Constraints from the computation of charges}
\label{appconstraintcharges}
\unboldmath
In this appendix, we compute the charges of \eqref{FOEOM} for the residual symmetries \eqref{KVBC}, \eqref{lamMaxBC} and \eqref{LambdaEOM}. The on-shell constraints do not directly lead to finite and integrable charges. 
The strategy we follow is to compute the variation of the charges and demand first that they are finite, and then integrable. 

We also know that we would like to keep the $\phi$-dependence of $Y$ in \eqref{KVBC}. 
It implies that the black string \eqref{BS4}, and the black string plus finite diffeomorphisms generated by $\xi(Y(u,\phi)=F(\phi))$ have to be included in the phase space. The latter is given by, in terms of the asymptotic ansatz, 
\begin{subequations}\label{BSplusdiffeo}
	\begin{align}\label{f00U00}
	& f_{00}(\phi)=\frac12 \log(\zeta \, F'(\phi) ) \,,\quad  U_{00}=\frac{  F''(\phi )}{\zeta \, F'(\phi )^3}\,,\quad  U_{01}=2\omega\,,\quad \Omega=\frac{2\omega}{L^2}\\
	&\beta_{00}(\phi)=-2 \log(\zeta \, F'(\phi) )\,,\quad  \alpha_{01}= \frac{4 L\, F'(\phi)}{ \sqrt{k_g} } \sqrt{- \omega^2 \,\zeta ^2+ r_- r_+}\\
	& V_{00}(u,\phi)=\frac{4}{ L F'(\phi )} \left( (r_++r_-) F'(\phi)^2 - \omega  \frac{F''(\phi )}{F'(\phi)} \right)
	\,,\quad  \bar V_{01}=-\frac4L-\frac{F''(\phi )^2}{L\,\zeta^2 \,F'(\phi )^4}
	\end{align}  
\end{subequations}
and all the other functions are put to zero.

We decide to start by examining the constraints coming from the Killing and Maxwell gauge parameters, \emph{i.e.} 
\begin{equation}\label{key}
k_{(\xi,\lambda,0)}(\Psi,\delta \Psi)\,,
\end{equation}
where $\xi$ is given by \eqref{KVBC}, $\lambda$ by \eqref{lamMaxBC}, $\Psi$ the asymptotically on-shell field configuration and $\delta \Psi$ its variation. 

As we will ultimately integrate over the surface at infinity, we only compute the component $ur$ of $k$ \eqref{kT}. 
The highest order we get is in $r$. However, using the EOM for $U_{00}$, it turns out to be a total derivative with respect to $\phi$ and thus will give a zero contribution to the charge, 
\begin{align}
&\frac1{16\pi }  \p_\phi \Big[ \Big(\p_p b_0  (-U_{00})+4 \left(2\p_p f_{00} \p_\phi  f_{00} -\p_\phi \p_p f_{00} \right) \text{e}^{b_0-4 f_{00}}+\p_p  U_{00}  \Big)X - Y \, \p_p b_0
\Big]
\,,
\end{align}
where the derivative with respect to $p$ represents the variation $\delta $ of the field content. 

The next order is in $\log(r/L)^3$ which is again a total derivative, 
\begin{align}
&\frac1{24\pi} \p_\phi \Big[ X\, \text{e}^{\beta_{00}} \left(2\p_\phi f_{00}\p_p f_{11} +4\p_p f_{00}\p_\phi f_{11} +\left(2 \p_\phi \p_p f_{00} -8\p_p f_{00} \p_\phi  f_{00} \right) f_{11} -\p_\phi \p_p f_{11} \right)\Big]\,. 
\end{align}
The next order goes like $\log(r/L)^2$. The treatment of this term is tedious but we have noticed that only the function $Y$ is present and not its derivative. It means that it is not rearrangeable as a total derivative, as we chose to discard the case where $Y$ is independent of $\phi$. 
The term proportional to $Y$ is 
\begin{align*}
& -\frac{\delta p \,k_g ^2\, \text{e}^{-8 f_{00}}}{512 \pi L^4 } 
\Big[ 
\p_p a_{00} \left( \p_\phi a_{11} - 4 a_{11 }\p_\phi  f_{00} \right)
a_{00}^2\\
& +  \Big(
-4\p_p a_{11}\p_\phi f_{00} +\p_pb_0  \left( \p_\phi a_{11} - 4 a_{11 }\p_\phi  f_{00} \right) -4\p_pf_{00}  \left( \p_\phi a_{11} - 4 a_{11 }\p_\phi  f_{00} \right) \\& \qquad +  \p_p\p_\phi a_{11} - 4 a_{11 }\p_p\p_\phi  f_{00} 
\Big)
a_{00}^3
\Big] +\frac{\delta p}{32 \pi \,L^2} \p_p \Big[ \text{e}^{-b_0} \Big( k_g \, \alpha_{21}a_{00} +2 \text{e}^{4f_{00}} U_{21}\,L^2 \Big)\Big]\,.
\end{align*}
One choice to cancel this term is to take
\begin{equation}\label{L2implic}
a_{00}=0 \,, \quad  U_{21}=0 
\end{equation}
However, $U_{21}$ is given by \eqref{EOMU} and so we have that $f_{11}=F_{11} (u)\text{e}^{2f_{00}} $.
Moreover, \eqref{EOMAu} implies that 
\begin{equation}\label{key}
\alpha_{00}=\alpha_{0}(u)
\end{equation}
and therefore 
\begin{equation}\label{key}
\lambda=\lambda(u)\,. 
\end{equation}

Now we consider the rest of the $\log(r/L)^2$ term. It is 
\begin{align}
\frac{\delta p}{32 \pi\,L^2} \Big[ &\p_p\Big( k_g\,\text{e}^{-b_0} \alpha_{21}\,\lambda \Big) + \p_p\Big(  X(u) \, k_g \,\text{e}^{-b_0} \alpha_{0}\, \alpha_{21} \Big)  \\
&  + X ( -2 V_{20} \p_p b_0 +2\p_p V_{20})  L^3 \Big]
\end{align} 
The function $\alpha_{21}$, given by \eqref{EOMAu}, is a total derivative with respect to $\phi$, as it is the case for $V_{20}$, given by \eqref{EOMV}, once the conditions \eqref{L2implic} are imposed. So, the divergence in $\log(r/L)^2$ is a total derivative with respect to $\phi$. 

Now we consider the divergent term in $\log(r/L)$. The $\lambda$ sector is a total derivative with respect to $\phi$ once we use the explicit form of $\alpha_{21}$ and $\alpha_{11}$. 
We consider the $Y$-sector and we realize that we can only recast one part of the $Y$-contribution into a total derivative. The remaining is 
\begin{equation}\label{key}
-\frac{\delta p \,Y}{4\pi} \left( 
\p_p( -2f_{01}\p_\phi f_{00}+\p_\phi f_{01}) +  F_{11}  \p_p \p_\phi \text{e}^{2f_{00}}
\right) \,,
\end{equation}
and the $X $-sector is 
\begin{align*}
&	\frac{\delta p}{2\pi}X   \p_p f_{00}\Big[
\text{e}^{-2 f_{00}+b_0}  \left( 12(\p_\phi f_{00})^2-4\p_\phi^2 f_{00}\right)+ \text{e}^{2f_{00}} \left(4\p_\phi f_{00} U_{00}-4 \p_u f_{00}+3 \p_\phi U_{00}-4 \text{e}^{b_0}\right)\Big]F_{11}\\
&+	\frac{\delta p}{4 \pi}X   \p_p f_{00}\Big[
\text{e}^{2 f_{00}} \left(2\p_\phi f_{00} U_{00}-2 \p_u f_{00}+\p_\phi U_{00}\right)
\p_pF_{11} 
- 2\text{e}^{2 f_{00}} \p_p f_{00}\p_uF_{11}	\Big]\\
& + \frac{\delta p\,k_g }{16\pi L^2}X \alpha_{0}  \Big( 2 \p_\phi( -\text{e}^{-4f_{00}}a_{01} )+\p_p(\text{e}^{-b_0 }\alpha_{21} )  \Big)\\
&	 +	\frac{\delta p}{16\pi}X  
\left(\text{e}^{4 f_{00}-b_0}  U_{00} \left(\p_p b_0-4 \p_p f_{00}\right)+ 8 \p_p f_{00} \p_\phi f_{00}\right)U_{11}\\
&+	\frac{\delta p}{16\pi}	X  \Big[ 
\p_p b_0 V_{10}\,L- U_{00} \p_p U_{11}  \text{e}^{4 f_{00}-b_0}+4 \p_p f_{00} \p_\phi U_{11}  -\p_p V_{10}\,L+ \p_\phi \p_p U_{11} -\p_pb_0 \p_\phi U_{11}
\Big]\,. 
\end{align*}
The choice
\begin{equation}\label{key}
F_{11}=0\,, \quad f_{01}=F_{01}(u)\text{e}^{2f_{00}}\,
\end{equation}
makes the $Y$ and $X$-sectors finite (recalling that $U_{11}$ is given by \eqref{EOMU} and $V_{10}$ is now a total derivative with respect to $\phi$). 

We have restricted our ansatz to ensure that the variation of the charges are finite. Before tackling the integrability question, we go back to the equations of motions. The dilaton equation at order $1/r^3$ now reads
\begin{equation}\label{key}
8 \p_u F_{01} \text{e}^{6 f_{00}-b_0}=0
\end{equation}
and forces us to take $F_{01}$ constant. 

With these new inputs, we consider the integrability of charges. 
The Maxwell part is integrable, 
\begin{equation}\label{key}
\delta H_{(0,\lambda,0)}=\frac{k_g}{32\pi\,L^2} \lambda \int \extd \phi \, \p_p \left( \text{e}^{-b_0} (\alpha_{01}-a_{01} \text{e}^{4 f_{00}}U_{00} )
\right) \,.
\end{equation}
However, only a part of the Virasoro sector is integrable,
\begin{equation}\label{Vircharge}
\delta H_{(Y,0,0)}=\frac{1}{16\pi}  \int \extd\phi \, \p_p \left( \text{e}^{-b_0+4f_{00}} U_{01} \, Y \right) \,, 
\end{equation}
and the remaining term $-\frac1{2\pi} \text{e}^{2f_{00}} F_{01} \p_p f_{00} \p_\phi Y $ is in general not integrable with respect to $p$ neither a total derivative with respect to $\phi$.
To make it a total derivative, we take $F_{01}$ independent of $p$. Thus it has to be the same for any point in the phase space, in particular for the black string, which has $F_{01}=0$. 
So the total Virasoro charge is given by \eqref{Vircharge}.

Now, we turn to the $X $-sector. First, we extract obvious total derivatives and directly integrable terms,  
\begin{equation}\label{key}
\frac1{16\pi}\left[ \p_p V_{00} \, L+ \p_\phi\left( \p_p b_0 \, U_{01} -\p_p U_{01} \right) 
\right] \,. 
\end{equation}
Also, there is only one other term involving $V_{00}$, namely $- \p_p b_0 \, V_{00}$. As $V_{00}$ is non zero for the solution \eqref{BSplusdiffeo}, we have to take $b_0$ independent of $p$ and so $b_0=0$, which is consistent with the black string. 
The terms involving $\alpha_0$ function are 
\begin{equation}\label{key}
\frac{k_g \, \delta p}{32\pi\,L^2}\Big[ \p_\phi\p_p\left(  - \text{e}^{-4f_{00}} a_{01} \alpha_0 \right) + \p_p \left( -a_{01} U_{00}+ \alpha_{01} \right) \alpha_{0}\Big] \,. 
\end{equation}
The first term is a total derivative with respect to $\phi$ while the second requires $\alpha_{0}$ to be independent of $p$ and thus is put to zero. Therefore, the gauge parameter of the Maxwell field becomes a constant $\lambda$. 
The last terms of the $X $-sector are 
\begin{align}\label{eq67}
&\frac{\delta p} {16\pi}\Big[  U_{00} \text{e}^{4  f_{00}} \left(\p_p U_{01}+ 4 U_{01}\p_p f_{00} \right)-4 \p_p f_{00}  \left(2 \p_\phi f_{00}  U_{01}+\p_\phi  U_{01}\right)\Big] \,,
\end{align}
which is equivalent to 
\begin{align}\label{eq67}
&\frac{\delta p} {16\pi}\Big[ - U_{01} (\text{e}^{4  f_{00}} \p_p U_{00}+ 8  \p_\phi f_{00} \p_p f_{00} - 4 \p_p \p_\phi f_{00} ) + \p_\phi ( -4\p_p f_{00} U_{01} ) +\p_p(\text{e}^{4f_{00}} U_{00} U_{01} )\Big] \,. 
\end{align}
We need to make the first term a total derivative or to cancel it. Because of the solution \eqref{BSplusdiffeo}, $U_{01}$ cannot be zero and has to depend on $p$. Thus the braket in the first term is either zero or it is a total derivative with respect to $\phi$ along with $U_{01}$ being independent of $\phi$. 
Moreover, the function $U_{00}$ takes the form \eqref{EOMU} 
\begin{equation}
U_{00}=\text{e}^{-2f_{00}} G(u,\phi) + \text{e}^{-4f_{00}}  \p_\phi f_{00} \,\quad \text{with } G(u,\phi)=\int^\phi \p_u \text{e}^{2f_{00}} +U_0\,.
\end{equation}
So the first braket in \eqref{eq67} becomes 
\begin{equation}
\text{e}^{2f_{00}} (-2 G \, \p_p f_{00}+ \p_p G) - 2 \p_p\p_\phi f_{00} \,.
\end{equation}
The first term is neither a total derivative with respect to $\phi$ neither can compensate the second term so it has to vanish, $G=0$. Thus, $U_0=0$ and 
\begin{equation}\label{f00phi}
f_{00}=f_{00}(\phi) \,.
\end{equation}
The second term is a total derivative with respect to $\phi$, which requires $U_{01}$ to be independent of $\phi$. Moreover, with the new constraints, the equation of motion for $U_{01}$ becomes $\p_u U_{01}=0$. Thus, $U_{01}$ is a constant which is still consistent with \eqref{BSplusdiffeo}. 

Also, the equation \eqref{f00phi} implies that the $Y$ function is now only a function of $\phi$, to still preserve the asymptotic fall-off of the dilaton. Moreover, studying the preservation of the $\phi$-dependence of the $uu$-component of the metric forces us to take the function $X(u)$ to be constant $X$. 

Finally, the charge associated to $X$ is integrable and given by 
\begin{equation}\label{key}
H_{(X,0,0 )}=\frac {X\, L}{16\pi}\int_0^{2\pi} \extd \phi\, V_{00} \,. 
\end{equation}

We have restrained our set of boundary conditions such that the charges associated to $\xi$ and $\lambda$ are finite and integrable. Now, we consider the last charge associated to $\Lambda$. It turns out that the charge is finite and integrable and that only the $\Lambda=\Lambda_\phi \extd \phi$ leads to a non zero charge, 
\begin{equation}\label{key}
\delta H_{(0,0,\Lambda_\phi \extd\phi)}=\frac1{16\pi}\int_0^{2\pi} \extd \phi\,  \frac{\p_p \Omega}{L^2}\Lambda_\phi\,. 
\end{equation}

In conclusion, by making choices on some of the arbitrary functions in \eqref{FOEOM}, we have managed to reach a phase space with finite and integrable charges. 

\boldmath
\section{The Horne--Horowitz black string}\label{appHH}
\label{}
\unboldmath


In this appendix, we explicit the coordinates transformation to reach Bondi gauge for the Horne--Horowitz black string \cite{Horne:1991gn} consisting in the neutral black string \eqref{BS4}, obtained by setting $\alpha=0$ .

In the Einstein frame, the action considered by Horne--Horowitz is 
\begin{equation}\label{key}
I=\frac{1}{16\pi G_3} \int \text{d}^3x\sqrt{-g} \left(
R+(\nabla\Phi_{\text{H}})^2  -\frac1{12} H^2 \text{e}^{ 4\Phi_{\text{H}}} +\frac 8k \text{e}^{-2\Phi_{\text{H}}}
 \right) \,.
\end{equation}
To adapt to our conventions, we take
\begin{equation}\label{HH-BS}
\Phi_{\text{H}} := - 2 \Phi \,,\qquad \frac 8k=\frac{4}{L^2 }\,.
\end{equation}
The non-extremal Horne--Horowitz black string takes the form 
\begin{align}
& \extd s_{\text{H}}^2=r (M_{\text{H}}-r)\,\extd t^2-\frac{ M_{\text{H}} r^2}{4 (M_{\text{H}}-r) \left(M_{\text{H}} r-Q_{\text{H}}^2\right)}\,\extd r^2+(r^2-\frac{Q_{\text{H}}^2 }{M_{\text{H}}}r)\,\extd x^2 \\
& \Phi _{\text{H}}=\log \left(\frac rL\right) \\
&(H_{trx})_{\text{H}}=-\frac{ L^2 \, Q_{\text{H}}}{ r^2}\,,
\end{align}	
where $|Q|<M$. 

We reach Bondi gauge with the following coordinate transformation
\begin{align}\nonumber
&u=-\frac{1}{4} \log \left( \frac{(r -M_{\text{H}})^{\frac{M_{\text{H}}^2}{M_{\text{H}}^2-Q_{\text{H}}^2}} }{ \left(r -\frac{Q_{\text{H}}^2}{M_{\text{H}}}\right)^{\frac{Q_{\text{H}}^2}{M_{\text{H}}^2-Q_{\text{H}}^2}}} \right)-\frac12\frac{M_{\text{H}} t-Q_{\text{H}} x}{\sqrt{M_{\text{H}}^2-Q_{\text{H}}^2}}\\\label{HHBondi}
&\phi=\frac{M_{\text{H}} Q_{\text{H}} \log \left(\frac{r -M_{\text{H}}}{r -\frac{Q_{\text{H}}^2}{M_{\text{H}}}}\right)}{2 \left(M_{\text{H}}^2-Q_{\text{H}}^2\right)}+\frac{Q_{\text{H}} t-M_{\text{H}} x}{\sqrt{M_{\text{H}}^2-Q_{\text{H}}^2}}\,,
\end{align}
in which the Horne--Horowitz black string takes the form
\begin{align}
& \extd s_{\text{H}}^2=4\left (-r^2+r\left(M_{\text{H}}+\frac{Q_{\text{H}}^2}{M_{\text{H}}}\right) \right )\extd u^2-2 r\,\extd u\,\extd r+4Q_{\text{H}}\,r\,\extd u\,\extd \phi+r^2\,\extd \phi^2 \\
& \Phi_{\text{H}} =\log \left(\frac rL\right) \\
&(H_{ur\phi})_{\text{H}}=\frac{ L^2 \, Q_{\text{H}}}{ r^2}\,.
\end{align}	

It is included in our phase space \eqref{BC} and corresponds to the case: 
\begin{equation}\label{key}
f_{00}=0\,, \quad u_{01}=2Q_{\text{H}}\,, \quad  V_{00}=\frac4L\left( M_{\text{H}}+\frac{Q_{\text{H}}^2}{M_{\text{H}}} \right) \,, \quad \Omega=\frac{2Q_{\text{H}}}{L^2}
\end{equation}
with the other functions and subleadings put to zero.
It is an extremal case from the point of view of the black string obtained by taking
\begin{equation}\label{key}
\zeta=1 \,, \quad \omega=Q_{\text{H}} \,,\quad r_+=M_{\text{H}}\,,\quad r_-=\frac{Q^2_{\text{H}}}{M_{\text{H}}}  \,.
\end{equation}
Thus, the thermodynamic considerations also hold for the Horne--Horowitz black string in Bondi gauge. 
However, an important remark is that the change of coordinates \eqref{HHBondi} is Horne--Horowitz parameters dependant and so we cannot pretend to describe the thermodynamic properties in the original system of coordinates.



\begin{thebibliography}{10}

\bibitem{Brown:1986nw}
J.~D. Brown and M.~Henneaux, ``{Central charges in the canonical realization of
  asymptotic symmetries: an example from three-dimensional gravity},'' {\em
  Commun. Math. Phys.} {\bf 104} (1986)
207.

\bibitem{Banados:1992wn}
M.~Ba\~nados, C.~Teitelboim, and J.~Zanelli, ``{The black hole in
  three-dimensional spacetime},'' {\em Phys. Rev. Lett.} {\bf 69} (1992)
  1849,
\href{http://www.arXiv.org/abs/hep-th/9204099}{{\tt hep-th/9204099}}.

\bibitem{Banados:1992gq}
M.~Ba\~nados, M.~Henneaux, C.~Teitelboim, and J.~Zanelli, ``{Geometry of the
  $2+1$ black hole},'' {\em Phys. Rev.} {\bf D48} (1993) 1506,
\href{http://www.arXiv.org/abs/gr-qc/9302012}{{\tt gr-qc/9302012}}.

\bibitem{Strominger:1997eq}
A.~Strominger, ``{Black hole entropy from near horizon microstates},'' {\em
  JHEP} {\bf 9802} (1998) 009,
\href{http://www.arXiv.org/abs/hep-th/9712251}{{\tt hep-th/9712251}}.

\bibitem{Witten:2007kt}
E.~Witten, ``{Three-dimensional gravity revisited},''
\href{http://www.arXiv.org/abs/0706.3359}{{\tt 0706.3359}}.

\bibitem{Maloney:2007ud}
A.~Maloney and E.~Witten, ``{Quantum gravity partition functions in three
  dimensions},'' {\em JHEP} {\bf 1002} (2010) 029,
\href{http://www.arXiv.org/abs/0712.0155}{{\tt 0712.0155}}.

\bibitem{Gaberdiel:2007ve}
M.~R. Gaberdiel, ``{Constraints on extremal self-dual CFTs},'' {\em JHEP} {\bf
  11} (2007) 087,
\href{http://www.arXiv.org/abs/0707.4073}{{\tt 0707.4073}}.

\bibitem{Keller:2011xi}
C.~A. Keller, ``{Phase transitions in symmetric orbifold CFTs and
  universality},'' {\em JHEP} {\bf 1103} (2011) 114,
\href{http://www.arXiv.org/abs/1101.4937}{{\tt 1101.4937}}.

\bibitem{Hartman:2014oaa}
T.~Hartman, C.~A. Keller, and B.~Stoica, ``{Universal spectrum of two-dimensional conformal
 field theory in the large-$c$ limit},'' {\em JHEP} {\bf 1409} (2014) 118,
\href{http://www.arXiv.org/abs/1405.5137}{{\tt 1405.5137}}.

\bibitem{Bondi:1962px}
H.~Bondi, M.~G.~J. van~der Burg, and A.~W.~K. Metzner, ``{Gravitational waves
  in general relativity. 7. Waves from axisymmetric isolated systems},'' {\em
  Proc. Roy. Soc. Lond.} {\bf A269} (1962)
21.

\bibitem{Sachs:1962wk}
R.~K. Sachs, ``{Gravitational waves in general relativity. 8. Waves in
  asymptotically flat spacetimes},'' {\em Proc. Roy. Soc. Lond.} {\bf A270}
  (1962)
103.

\bibitem{Sachs:1962zza}
R.~Sachs, ``{Asymptotic symmetries in gravitational theory},'' {\em Phys. Rev.}
  {\bf 128} (1962)
2851.

\bibitem{Strominger:2017zoo}
A.~Strominger, ``{Lectures on the infrared structure of gravity and gauge
  theory},''
\href{http://www.arXiv.org/abs/1703.05448}{{\tt 1703.05448}}.

\bibitem{Hawking:2016msc}
S.~W. Hawking, M.~J. Perry, and A.~Strominger, ``{Soft hair on black holes},''
  {\em Phys. Rev. Lett.} {\bf 116} (2016) 231301,
\href{http://www.arXiv.org/abs/1601.00921}{{\tt 1601.00921}}.

\bibitem{Barnich:2017ubf}
G.~Barnich, ``{Centrally extended BMS$_4$ Lie algebroid},'' {\em JHEP} {\bf 06}
  (2017) 007,
\href{http://www.arXiv.org/abs/1703.08704}{{\tt 1703.08704}}.

\bibitem{Compere:2013bya}
G.~Comp\`ere, W.~Song, and A.~Strominger, ``{New boundary conditions for AdS$_3$},''
  {\em JHEP} {\bf 05} (2013) 152,
\href{http://www.arXiv.org/abs/1303.2662}{{\tt 1303.2662}}.

\bibitem{Troessaert:2013fma}
C.~Troessaert, ``{Enhanced asymptotic symmetry algebra of AdS$_{3}$},'' {\em
  JHEP} {\bf 08} (2013) 044,
\href{http://www.arXiv.org/abs/1303.3296}{{\tt 1303.3296}}.

\bibitem{Avery:201three-dimensionalja}
S.~G. Avery, R.~R. Poojary, and N.~V. Suryanarayana, ``{An sl(2,$\mathbb{R}$)
  current algebra from AdS$_3$ gravity},'' {\em JHEP} {\bf 01} (2014) 144,
\href{http://www.arXiv.org/abs/1304.4252}{{\tt 1304.4252}}.

\bibitem{Grumiller:2016pq}
D.~Grumiller and M.~Riegler, ``{Most general AdS$_{3}$ boundary conditions},''
  {\em JHEP} {\bf 10} (2016) 023,
\href{http://www.arXiv.org/abs/1608.01308}{{\tt 1608.01308}}.

\bibitem{Perez:2016vqo}
A.~P\'erez, D.~Tempo, and R.~Troncoso, ``{Boundary conditions for general
  relativity on AdS$_{3}$ and the KdV hierarchy},'' {\em JHEP} {\bf 06} (2016)
  103,
\href{http://www.arXiv.org/abs/1605.04490}{{\tt 1605.04490}}.

\bibitem{Donnay:2015abr}
L.~Donnay, G.~Giribet, H.~A. Gonzalez, and M.~Pino, ``{Supertranslations and
  superrotations at the black hole horizon},'' {\em Phys. Rev. Lett.} {\bf 116}
  (2016) 091101,
\href{http://www.arXiv.org/abs/1511.08687}{{\tt 1511.08687}}.

\bibitem{Afshar:2015wjm}
H.~Afshar, S.~Detournay, D.~Grumiller, and B.~Oblak, ``{Near-horizon geometry
  and warped conformal symmetry},'' {\em JHEP} {\bf 03} (2016) 187,
\href{http://www.arXiv.org/abs/1512.08233}{{\tt 1512.08233}}.

\bibitem{Afshar:2016wfy}
H.~Afshar, S.~Detournay, D.~Grumiller, W.~Merbis, A.~Perez, D.~Tempo, and
  R.~Troncoso, ``{Soft Heisenberg hair on black holes in three dimensions},''
  {\em Phys. Rev.} {\bf D93} (2016) 101503,
\href{http://www.arXiv.org/abs/1603.04824}{{\tt 1603.04824}}.

\bibitem{Anninos:2011vd}
D.~Anninos, S.~de~Buyl, and S.~Detournay, ``{Holography for a de Sitter-Esque
 geometry},'' {\em JHEP} {\bf 05} (2011) 003,
\href{http://www.arXiv.org/abs/1102.3178}{{\tt 1102.3178}}.

\bibitem{Barnich:2006av}
G.~Barnich and G.~Comp\`ere, ``{Classical central extension for asymptotic
  symmetries at null infinity in three spacetime dimensions},'' {\em
  Class. Quant. Grav.} {\bf 24} (2007) F15,
\href{http://www.arXiv.org/abs/gr-qc/0610130}{{\tt gr-qc/0610130}}.

\bibitem{Detournay:2016sfv}
S.~Detournay and M.~Riegler, ``{Enhanced asymptotic symmetry algebra of $2+1$
  dimensional flat space},'' {\em Phys. Rev.} {\bf D95} (2017) 046008,
\href{http://www.arXiv.org/abs/1612.00278}{{\tt 1612.00278}}.

\bibitem{Grumiller:2017sjh}
D.~Grumiller, W.~Merbis, and M.~Riegler, ``{Most general flat space boundary
  conditions in three-dimensional Einstein gravity},'' {\em Class. Quant.
  Grav.} {\bf 34} (2017) 184001,
\href{http://www.arXiv.org/abs/1704.07419}{{\tt 1704.07419}}.

\bibitem{Ida:2000jh}
D.~Ida, ``{No black hole theorem in three-dimensional gravity},'' {\em Phys.
  Rev. Lett.} {\bf 85} (2000) 3758,
\href{http://www.arXiv.org/abs/gr-qc/0005129}{{\tt gr-qc/0005129}}.

\bibitem{Cornalba:2002fi}
L.~Cornalba and M.~S. Costa, ``{A new cosmological scenario in string
  theory},'' {\em Phys. Rev.} {\bf D66} (2002) 066001,
\href{http://www.arXiv.org/abs/hep-th/0203031}{{\tt hep-th/0203031}}.

\bibitem{Barnich:2012xq}
G.~Barnich, ``{Entropy of three-dimensional asymptotically flat cosmological
  solutions},'' {\em JHEP} {\bf 1210} (2012) 095,
\href{http://www.arXiv.org/abs/1208.4371}{{\tt 1208.4371}}.

\bibitem{Bagchi:2012xr}
A.~Bagchi, S.~Detournay, R.~Fareghbal, and J.~Simon, ``{Holography of three-dimensional flat
  cosmological horizons},'' {\em Phys. Rev. Lett.} {\bf 110} (2013) 141302,
\href{http://www.arXiv.org/abs/1208.4372}{{\tt 1208.4372}}.

\bibitem{Horne:1991gn}
J.~H. Horne and G.~T. Horowitz, ``{Exact black string solutions in
  three-dimensions},'' {\em Nucl. Phys.} {\bf B368} (1992) 444,
\href{http://www.arXiv.org/abs/hep-th/9108001}{{\tt hep-th/9108001}}.

\bibitem{Oliva:2009ip}
J.~Oliva, D.~Tempo, and R.~Troncoso, ``{Three-dimensional black holes,
  gravitational solitons, kinks and wormholes for BHT massive gravity},'' {\em
  JHEP} {\bf 07} (2009) 011,
\href{http://www.arXiv.org/abs/0905.1545}{{\tt 0905.1545}}.

\bibitem{Fareghbal:2014kfa}
R.~Fareghbal and S.~M. Hosseini, ``{Holography of three-dimensional asymptotically flat black
  holes},'' {\em Phys. Rev.} {\bf D91} (2015) 084025,
\href{http://www.arXiv.org/abs/1412.2569}{{\tt 1412.2569}}.

\bibitem{Carlip:2017xne}
S.~Carlip, ``{Black Hole Entropy from Bondi-Metzner-Sachs Symmetry at the
  Horizon},'' {\em Phys. Rev. Lett.} {\bf 120} (2018), no.~10, 101301,
\href{http://www.arXiv.org/abs/1702.04439}{{\tt 1702.04439}}.

\bibitem{Witten:1991yr}
E.~Witten, ``{On string theory and black holes},'' {\em Phys. Rev.} {\bf D44}
  (1991)
314.

\bibitem{Detournay:2005fz}
S.~Detournay, D.~Orlando, P.~M. Petropoulos, and P.~Spindel,
  ``{Three-dimensional black holes from deformed anti-de Sitter},'' {\em JHEP}
  {\bf 07} (2005) 072,
\href{http://www.arXiv.org/abs/hep-th/0504231}{{\tt hep-th/0504231}}.

\bibitem{Spindel:2018cgm}
P.~Spindel, ``{Three dimensional black strings: instabilities and asymptotic
  charges},''
\href{http://www.arXiv.org/abs/1810.00603}{{\tt 1810.00603}}.

\bibitem{Kaloper:1998vw}
N.~Kaloper, ``{Entropy count for extremal three-dimensional black strings},''
  {\em Phys. Lett.} {\bf B434} (1998) 285,
\href{http://www.arXiv.org/abs/hep-th/9804062}{{\tt hep-th/9804062}}.

\bibitem{Barnich:2015jua}
G.~Barnich, P.-H. Lambert, and P.~Mao, ``{Three-dimensional asymptotically flat
  Einstein--Maxwell theory},'' {\em Class. Quant. Grav.} {\bf 32} (2015) 245001,
\href{http://www.arXiv.org/abs/1503.00856}{{\tt 1503.00856}}.

\bibitem{Winicour85}
J.~Winicour, ``{Logarithmic asymptotic flatness},'' {\em Foundations of
  Physics} {\bf 15} (1985) 605.

\bibitem{Bagchi:2012yk}
A.~Bagchi, S.~Detournay, and D.~Grumiller, ``{Flat-space chiral gravity},''
  {\em Phys. Rev. Lett.} {\bf 109} (2012) 151301,
\href{http://www.arXiv.org/abs/1208.1658}{{\tt 1208.1658}}.

\bibitem{Banados:1998gg}
M.~Ba\~nados, ``{Three-dimensional quantum geometry and black holes},'' {\em AIP
  Conf. Proc.} {\bf 484} (1999) 147,
\href{http://www.arXiv.org/abs/hep-th/9901148}{{\tt hep-th/9901148}}.

\bibitem{Iyer:1994ys}
V.~Iyer and R.~M. Wald, ``{Some properties of N\oe ther charge and a proposal for
  dynamical black hole entropy},'' {\em Phys. Rev.} {\bf D50} (1994) 846,
\href{http://www.arXiv.org/abs/gr-qc/9403028}{{\tt gr-qc/9403028}}.

\bibitem{Abbott:1982jh}
L.~F. Abbott and S.~Deser, ``{Charge definition in nonabelian gauge
 theories},'' {\em Phys. Lett.} {\bf 116B} (1982)
259.

\bibitem{Anderson:1996sc}
I.~M. Anderson and C.~G. Torre, ``{Asymptotic conservation laws in field
  theory},'' {\em Phys. Rev. Lett.} {\bf 77} (1996) 4109,
\href{http://www.arXiv.org/abs/hep-th/9608008}{{\tt hep-th/9608008}}.

\bibitem{Barnich:2005kq}
G.~Barnich and G.~Comp\`ere, ``{Conserved charges and thermodynamics of the
  spinning G\"odel black hole},'' {\em Phys. Rev. Lett.} {\bf 95} (2005) 031302,
\href{http://www.arXiv.org/abs/hep-th/0501102}{{\tt hep-th/0501102}}.

\bibitem{Compere:2007vx}
G.~Comp\`ere, ``{Note on the first law with p-form potentials},'' {\em Phys.
  Rev.} {\bf D75} (2007) 124020,
\href{http://www.arXiv.org/abs/hep-th/0703004}{{\tt hep-th/0703004}}.

\bibitem{Detournay:201two-dimensionalz}
S.~Detournay and M.~Guica, ``{Stringy Schr\"odinger truncations},'' {\em JHEP}
  {\bf 1308} (2013) 121,
\href{http://www.arXiv.org/abs/1212.6792}{{\tt 1212.6792}}.

\bibitem{Barnich:2001jy}
G.~Barnich and F.~Brandt, ``{Covariant theory of asymptotic symmetries,
  conservation laws and central charges},'' {\em Nucl. Phys.} {\bf B633} (2002)
  3,
\href{http://www.arXiv.org/abs/hep-th/0111246}{{\tt hep-th/0111246}}.

\bibitem{Barnich:2007bf}
G.~Barnich and G.~Comp\`ere, ``{Surface charge algebra in gauge theories and
  thermodynamic integrability},'' {\em J. Math. Phys.} {\bf 49} (2008) 042901,
\href{http://www.arXiv.org/abs/0708.2378}{{\tt 0708.2378}}.

\bibitem{Chen:2013aza}
B.~Chen, J.-j. Zhang, J.-d. Zhang, and D.-l. Zhong, ``{Aspects of warped
  AdS$_3$/CFT$_2$ correspondence},'' {\em JHEP} {\bf 1304} (2013) 055,
\href{http://www.arXiv.org/abs/1302.6643}{{\tt 1302.6643}}.

\bibitem{Compere:2018aar}
G.~Comp\`ere and A.~Fiorucci, ``{Advanced lectures in general relativity},''
\href{http://www.arXiv.org/abs/1801.07064}{{\tt 1801.07064}}.

\bibitem{Wald:1999wa}
R.~M. Wald and A.~Zoupas, ``{A General definition of ``conserved quantities'' in
  general relativity and other theories of gravity},'' {\em Phys. Rev.} {\bf
  D61} (2000) 084027,
\href{http://www.arXiv.org/abs/gr-qc/9911095}{{\tt gr-qc/9911095}}.

\bibitem{Barnich:2011mi}
G.~Barnich and C.~Troessaert, ``{BMS charge algebra},'' {\em JHEP} {\bf 12}
  (2011) 105,
\href{http://www.arXiv.org/abs/1106.0213}{{\tt 1106.0213}}.

\bibitem{Compere:2007in}
G.~Comp\`ere and S.~Detournay, ``{Centrally extended symmetry algebra of
  asymptotically G\"odel spacetimes},'' {\em JHEP} {\bf 0703} (2007) 098,
\href{http://www.arXiv.org/abs/hep-th/0701039}{{\tt hep-th/0701039}}.

\bibitem{BS3}
I.~Bars and K.~Sfetsos, ``{Global analysis of new gravitational singularities
  in string and particle theories},'' {\em Phys. Rev.} {\bf D46} (1992)
  4495,
\href{http://www.arXiv.org/abs/hep-th/9205037}{{\tt hep-th/9205037}}.

\bibitem{Bars:1992sr}
I.~Bars and K.~Sfetsos, ``{Conformally exact metric and dilaton in string
  theory on curved spacetime},'' {\em Phys. Rev.} {\bf D46} (1992) 4510,
\href{http://www.arXiv.org/abs/hep-th/9206006}{{\tt hep-th/9206006}}.

\bibitem{BCR}
K.~Bardakci, M.~J. Crescimanno, and E.~Rabinovici, ``{Parafermions from coset
  models},'' {\em Nucl. Phys.} {\bf B344} (1990)
344.

\bibitem{parafZF}
V.~A. Fateev and A.~B. Zamolodchikov, ``{Parafermionic currents in the
  two-dimensional conformal quantum field theory and selfdual critical points
  in $Z_n$ invariant statistical systems},'' {\em Sov. Phys. JETP} {\bf 62}
  (1985) 215.
[Zh. Eksp. Teor. Fiz. 89, 380(1985)].

\bibitem{lykken}
J.~D. Lykken, ``{Finitely reducible realizations of the $N=2$ superconformal
  algebra},'' {\em Nucl. Phys.} {\bf B313} (1989)
473.

\bibitem{PS05}
P.~Marios~Petropoulos and K.~Sfetsos, ``{NS5-branes on an ellipsis and novel
  marginal deformations with parafermions},'' {\em JHEP} {\bf 01} (2006) 167,
\href{http://www.arXiv.org/abs/hep-th/0512251}{{\tt hep-th/0512251}}.

\bibitem{PS06}
P.~M. Petropoulos and K.~Sfetsos, ``{Non-Abelian coset string backgrounds from
  asymptotic and initial data},'' {\em JHEP} {\bf 04} (2007) 033,
\href{http://www.arXiv.org/abs/hep-th/0610055}{{\tt hep-th/0610055}}.

\bibitem{Kiritsis:1995iu}
E.~Kiritsis and C.~Kounnas, ``{Infrared behavior of closed superstrings in
  strong magnetic and gravitational fields},'' {\em Nucl. Phys.} {\bf B456}
  (1995) 699,
\href{http://www.arXiv.org/abs/hep-th/9508078}{{\tt hep-th/9508078}}.

\bibitem{Israel:2004vv}
D.~Isra\" el, C.~Kounnas, D.~Orlando, and P.~M. Petropoulos, ``{Electric/magnetic
  deformations of $S^3$ and AdS$_3$, and geometric cosets},'' {\em Fortsch. Phys.}
  {\bf 53} (2005) 73,
\href{http://www.arXiv.org/abs/hep-th/0405213}{{\tt hep-th/0405213}}.

\bibitem{Petropoulos:2005}
P.~M. Petropoulos, ``{Deformations and geometric cosets},'' in {\em {The
  quantum structure of spacetime and the geometric nature of fundamental
  interactions. Proceedings, 4th Meeting, RTN2004, Kolymbari, Crete, Greece,
  September 5-10, 2004,}}
\newblock
\href{http://www.arXiv.org/abs/hep-th/0412328}{{\tt hep-th/0412328}}.
\newblock

\bibitem{Israel:2004cd}
D.~Isra\" el, C.~Kounnas, D.~Orlando, and P.~M. Petropoulos, ``{Heterotic strings
  on homogeneous spaces},'' {\em Fortsch. Phys.} {\bf 53} (2005) 1030,
\href{http://www.arXiv.org/abs/hep-th/0412220}{{\tt hep-th/0412220}}.

\bibitem{Gao:2001ut}
S.~Gao and R.~M. Wald, ``{The ``Physical process'' version of the first law and
  the generalized second law for charged and rotating black holes},'' {\em
  Phys. Rev.} {\bf D64} (2001) 084020,
\href{http://www.arXiv.org/abs/gr-qc/0106071}{{\tt gr-qc/0106071}}.

\bibitem{Copsey:2005se}
K.~Copsey and G.~T. Horowitz, ``{The role of dipole charges in black hole
  thermodynamics},'' {\em Phys. Rev.} {\bf D73} (2006) 024015,
\href{http://www.arXiv.org/abs/hep-th/0505278}{{\tt hep-th/0505278}}.

\bibitem{Wald:1993nt}
R.~M. Wald, ``Black hole entropy is the N\oe ther charge,'' {\em Phys. Rev.} {\bf
  D48} (1993) 3427--3431,
\href{http://arXiv.org/abs/gr-qc/9307038}{{\tt gr-qc/9307038}}.

\bibitem{Compere:2006my}
G.~Comp\`ere, ``{An introduction to the mechanics of black holes},'' in {\em {2nd
  Modave Summer School in Theoretical Physics Modave, Belgium, August 6-12,
  2006,}}\newblock
\href{http://www.arXiv.org/abs/gr-qc/0611129}{{\tt gr-qc/0611129}}.
\newblock

\bibitem{Dolan:2014jva}
B.~P. Dolan, ``{Black holes and Boyle's law -- the thermodynamics of the
  cosmological constant},'' {\em Mod. Phys. Lett.} {\bf A30} (2015)
  1540002,
\href{http://www.arXiv.org/abs/1408.4023}{{\tt 1408.4023}}.

\bibitem{Caldarelli:1999xj}
M.~M. Caldarelli, G.~Cognola, and D.~Klemm, ``{Thermodynamics of
  Kerr--Newman-AdS black holes and conformal field theories},'' {\em Class.
  Quant. Grav.} {\bf 17} (2000) 399,
\href{http://www.arXiv.org/abs/hep-th/9908022}{{\tt hep-th/9908022}}.

\bibitem{Castro:2012av}
A.~Castro and M.~J. Rodriguez, ``{Universal properties and the first law of
  black hole inner mechanics},'' {\em Phys. Rev.} {\bf D86} (2012) 024008,
\href{http://www.arXiv.org/abs/1204.1284}{{\tt 1204.1284}}.

\bibitem{Chen:2012mh}
B.~Chen, S.-x. Liu, and J.-j. Zhang, ``{Thermodynamics of black hole horizons
  and Kerr/CFT correspondence},'' {\em JHEP} {\bf 1211} (2012) 017,
\href{http://www.arXiv.org/abs/1206.2015}{{\tt 1206.2015}}.

\bibitem{Larsen:1997ge}
F.~Larsen, ``{A string model of black hole microstates},'' {\em Phys. Rev.} {\bf
  D56} (1997) 1005,
\href{http://www.arXiv.org/abs/hep-th/9702153}{{\tt hep-th/9702153}}.

\bibitem{Detournay:2012ug}
S.~Detournay, ``{Inner mechanics of three-dimensional black holes},'' {\em Phys. Rev. Lett.}
  {\bf 109} (2012) 031101,
\href{http://www.arXiv.org/abs/1204.6088}{{\tt 1204.6088}}.

\bibitem{Hofman:2011zj}
D.~M. Hofman and A.~Strominger, ``{Chiral scale and conformal invariance in two-dimensional
 quantum field theory},'' {\em Phys. Rev. Lett.} {\bf 107} (2011) 161601,
\href{http://www.arXiv.org/abs/1107.2917}{{\tt 1107.2917}}.

\bibitem{Detournay:2012pc}
S.~Detournay, T.~Hartman, and D.~M. Hofman, ``{Warped conformal field
  theory},'' {\em Phys. Rev.} {\bf D86} (2012) 124018,
\href{http://www.arXiv.org/abs/1210.0539}{{\tt 1210.0539}}.

\bibitem{Dijkgraaf:1991ba}
R.~Dijkgraaf, H.~L. Verlinde, and E.~P. Verlinde, ``{String propagation in a
  black hole geometry},'' {\em Nucl. Phys.} {\bf B371} (1992)
269.

\bibitem{Sfetsos:1992yi}
K.~Sfetsos, ``{Conformally exact results for $SL(2,R) \times SO(1,1)^{d-2} / SO(1,1)$
  coset models},'' {\em Nucl. Phys.} {\bf B389} (1993) 424,
\href{http://www.arXiv.org/abs/hep-th/9206048}{{\tt hep-th/9206048}}.

\bibitem{Horowitz:1993jc}
G.~T. Horowitz and D.~L. Welch, ``{Exact three-dimensional black holes in
  string theory},'' {\em Phys. Rev. Lett.} {\bf 71} (1993) 328--331,
\href{http://www.arXiv.org/abs/hep-th/9302126}{{\tt hep-th/9302126}}.

\bibitem{Barnich:2013sxa}
G.~Barnich and P.-H. Lambert, ``{Einstein--Yang--Mills theory: asymptotic
  symmetries},'' {\em Phys. Rev.} {\bf D88} (2013) 103006,
\href{http://www.arXiv.org/abs/1310.2698}{{\tt 1310.2698}}.

\end{thebibliography}

\end{document}